\documentclass[12pt]{article}

\usepackage{epsf,epsfig}

\usepackage{a4wide}

\def\beq{\begin{eqnarray}}
\def\eeq{\end{eqnarray}}

\def\be{\begin{equation}}
\def\ee{\end{equation}}

\def\scetI{SCET$_{\rm I}${}}
\def\scetII{SCET$_{\rm II}${}}

\begin{document}
\thispagestyle{empty}

\begin{flushright}
  CERN-PH-TH/04-155\\
  SLAC-PUB-10604\\
ICHEP04-110993
\\
hep-ph/0408188

\end{flushright}

\vspace{\baselineskip}

\begin{center}
\vspace{0.5\baselineskip}

{\bf \Large 
Non-factorizable Contributions to $B \to \pi\pi$ Decays}

\vspace{4\baselineskip}
{\large 
  Thorsten Feldmann$^a$ and 
  Tobias Hurth$^{a,b,}$\footnote{Heisenberg Fellow}}
\\
\vspace{2em}

{  \it 
$^a$ CERN, Dept.\ of Physics, Theory Division, CH-1211 Geneva, Switzerland}
\\[0.3em]
{ \it $^b$ SLAC, Stanford University, Stanford, CA 94309, USA}

\vspace{3\baselineskip}

%
\textbf{Abstract}
\vspace{1em}

\parbox{0.9\textwidth}
{\small We investigate to what extent the experimental information
on $B \to \pi\pi$ branching fractions and CP asymmetries 
can be used to better understand the QCD dynamics in these decays. 
For this purpose we decompose the independent isospin amplitudes
into factorizable and non-factorizable contributions. The former
can be estimated within the framework of QCD factorization for
exclusive $B$\/ decays. The latter vanish in the heavy-quark
limit, $m_b \to \infty$, and are treated as unknown
hadronic parameters. We discuss at some length in which way  the 
non-factorizable contributions are treated in different theoretical
and phenomenological frameworks. We point out the
potential differences between the phenomenological treatment
of power-corrections in the ``BBNS approach'', and the
appearance of power-suppressed operators in soft-collinear
effective theory (SCET). On that basis we define a
handful of different (but generic) scenarios where 
the non-factorizable part of isospin
amplitudes is parametrized in terms of three or four unknowns,
which can be constrained by data.
We also give some short discussion on the implications of our analysis
 for $B \to \pi K$
decays.  In particular, since non-factorizable QCD effects in 
$B \to \pi \pi$ may be large, 
we cannot exclude sizeable non-factorizable effects,  which
violate $SU(3)_F$ flavour symmetry, or even isospin symmetry
(via long-distance QED effects).
This may help to explain certain puzzles 
in connection with isospin-violating observables in $B \to \pi K$
decays.}
\end{center}



\clearpage
\setcounter{page}{1}

\section{Introduction}
\label{sec:intro}

An important unresolved question in the theoretical
analysis of charmless non-leptonic $B$ decays is the
quantitative understanding of the non-perturbative
dynamics responsible for non-factorizable 
contributions to decay amplitudes.

In the heavy quark limit, $m_{b} \to \infty$ , a
factorization theorem 
\cite{Beneke:1999br} states that $B \to PP$ decay amplitudes 
(where $P$ stands for a light pseudoscalar meson) 
factorize into perturbatively calculable coefficient
functions $T_{\rm I,II}$ and universal hadronic quantities, namely
$B \to P$ transition form factors  $F^{B \to P}(m_P^2)$ 
and light-cone wave functions
for heavy and light mesons ($\phi_B$ and $\phi_\pi$).
Schematically, one has 
\beq
\label{QCDF}
  \langle PP| H_{\rm eff}|B\rangle
&=&
   F^{B \to P} \cdot T_{\rm I} \otimes \phi_P
+
  T_{\rm II} \otimes \phi_B \otimes \phi_P \otimes \phi_P
\nonumber \\[0.2em]
&& {} +
 \mbox{terms suppressed by $1/m_b$}
\eeq
where the symbol $\otimes$ represents convolution
with respect to the light-cone momentum fractions of
light quarks inside the mesons.
Without taking into account radiative
corrections, $T_{\rm I}$ depends on kinematic factors only,
and $T_{\rm II}$ vanishes. This approximation corresponds
to the ``naive'' factorization assumption.
At first order of the strong coupling constant
the factorization formula (\ref{QCDF}) has been shown to hold by
the explicit calculation of the $\alpha_s$ corrections to
naive factorization \cite{Beneke:1999br,Beneke:2003zv}.
In the following we refer to (\ref{QCDF}) in the heavy-quark limit as
``QCD factorization''.
Arguments towards an all-order proof have been given
in an effective theory framework in \cite{Chay:2003zp}
(see also \cite{Bauer:2004tj} for a recent discussion). 
Here,  the relevant momentum modes for $B \to PP$ decays
are determined by the momentum scaling of
the external particles and their possible interactions.
The QCD factorization formula follows from identifying
$T_{\rm I,II}$ as coefficient functions of operators
in the soft-collinear effective theory (SCET), where the constituents
from different hadrons appear to be decoupled.
An all-order proof, which considers the subtle effects
of endpoint divergences 
(see Section~2 in \cite{Beneke:2003pa} for a toy example,
and also \cite{Lange:2003pk}),
has not been worked out to the last detail. It
should follow a similar line of reasoning as for the 
somewhat simpler cases of $B \to \gamma$ and $B \to \pi$
form factors \cite{Lunghi:2002ju,Bosch:2003fc,Beneke:2003pa,Lange:2003pk}.

A major complication for phenomenology arises due
to the fact that at least some of the 
power corrections to (\ref{QCDF}) appear to be enhanced by
large numerical coefficients (these are proportional
to the quark condensate in QCD, for this reason these
terms are referred to as ``chirally enhanced'').
In the diagrammatic analysis, non-factorizable
contributions are identified from convolution integrals
that suffer from endpoint-divergences when one of the
parton energies vanishes. The authors of 
\cite{Beneke:1999br} 
parametrize the chirally enhanced contributions of this
type by an arbitrary complex number. The modulus of this number
is estimated from regularizing the endpoint divergences with a
finite energy cut-off. 
We emphasize that this is a model-dependent procedure
that aims to get a quantitative handle on terms that
are beyond the QCD factorization approach. 
We will refer to this approximation
and its phenomenological implications as the ``BBNS approach''.

Another well-known framework is to use approximate flavour
symmetries (isospin or $SU(3)_F$) to relate different
decay amplitudes and reduce the number of unknown hadronic
parameters \cite{Gronau:1990ka,Nir:1991cu}. This procedure is often 
combined with {so-called}
``plausible dynamical assumptions'' about
the importance of certain flavour topologies that can
be identified in the factorization approximation only. 
In particular, from 
a recent study in \cite{Buras:2004ub} along these lines, it has
been concluded that (a) non-factorizable effects in
$B \to \pi\pi$ are large, (b) certain $B \to \pi K$
observables point to possible new physics effects
in electroweak penguin contributions.

A comprehensive analysis of presently available
$B \to PP$ data leads to a somewhat milder conclusion
\cite{Charles:2004jd}.
It is found that, within the statistical uncertainties,
the ``anomalies'' in $B \to \pi K$  decays may still be considered as
being consistent with the SM. 
Furthermore, the non-factorizable effects
in $B \to \pi\pi$ can be more or less accounted for by
fitting the hadronic input parameters in the
BBNS approach to experimental data. Of course, in
this procedure, most of the predictive power of the
QCD factorization approach is lost. Still, the 
model-dependent parametrization of non-factorizable
effects in the BBNS framework turns out to result
in significant constraints when used as the basis
for a CKM fit.

The purpose of this article is to carefully examine
different (model-dependent) approaches
to quantify non-factorizable hadronic effects 
in charmless non-leptonic $B$ decays.
In view of the ultimate goal, namely
to extract independent information on CKM \mbox{parameters}
from non-leptonic $B$ decays, we think that it is
important to make sure that certain assumptions
about the size of flavour symmetry breaking, the
origin of strong phases etc.\ are clearly identified in order
not to induce an uncontrolled theoretical bias.

Our paper is organized as follows.
In the next section 
we will use the $B \to \pi\pi$ decays as a guideline, and 
decompose the independent isospin amplitudes into
factorizable and non-factorizable parts.
The factorizable contributions are estimated within
the QCD factorization approach to first order in the
strong coupling constant, and using default values for
hadronic input parameters. The non-factorizable contributions
are considered as unknown parameters, the size of which 
has to be taken from experimental data on branching fractions
and CP asymmetries. 
We will discuss the origin/interpretation
of different sources for non-factorizable effects within
the BBNS approach, SCET, and phenomenological \mbox{studies}
assuming large long-distance penguin contributions. 
From this we develop a handful of constrained scenarios 
that implement different generic features of 
such approaches. Constraining the parameters
for these  scenarios using $B \to \pi\pi$ data,
one may obtain quite different results on the
size of individual isospin amplitudes. In particular, we find
that in scenarios with large non-factorizable 
penguin or annihilation contributions, one may not exclude sizeable
corrections to isospin-violating observables which arise
from long-distance QED effects, and may be relevant to
explain the $B \to \pi K$ puzzles mentioned above.
A detailed numerical analysis of $B \to \pi K$ modes 
is postponed until new experimental data become available.

\section{Factorizable and non-factorizable contributions}

\subsection{Isospin decomposition of decay amplitudes}

The CKM elements entering the effective weak hamiltonian 
for $b$\/-quark decays in the Standard Model are defined as
\beq
  \lambda_i^{(q)} &=& V_{ib} V_{iq}^* \ .
\eeq
In particular, for $b \to d$ transitions we have
\beq
  \frac{\lambda_u^{(d)}}{\lambda_c^{(d)}} = - R_u \, e^{- i \gamma} \ ,
\eeq
and for $b \to s$ transitions we have
\beq
  \frac{\lambda_u^{(s)}}{\lambda_c^{(s)}} = 
  \tan^2\theta_C \, R_u \, e^{- i \gamma} 
\equiv \epsilon_{KM}  \, e^{- i \gamma}
\eeq
with  $\sin\theta_C \simeq 0.2266$ being the Cabbibo angle, and
$R_u = \sqrt{\tilde \rho^2+\tilde \eta^2}\simeq 0.405$ and 
$\gamma \simeq 62^\circ$
related to one side and one angle of the CKM-triangle
(all numbers from \cite{Charles:2004jd},
$R_u$ is sometimes denoted as $R_b$).
In the case of $b \to d$ decays, all CKM factors are of
the same order, whereas for $b \to s $ decays,  $\epsilon_{KM}$
is suppressed by two powers of the Cabbibo angle.

Assuming isospin conservation in hadronic matrix elements
the $B \to \pi\pi$ decay amplitudes can be decomposed into\footnote{The 
$|\pi^0\pi^0\rangle$ state in this notation already
 includes a statistical factor $1/\sqrt2$ from Bose symmetry, i.e.\
 the branching ratio calculated with this amplitude does not 
 receive an additional factor $1/2$. This corresponds to 
 the convention in \cite{Buras:2004ub} and differs from the
 convention in \cite{Beneke:1999br}.}
\beq
 \sqrt2 \,    \langle \pi^- \pi^0 
    | H_{\rm eff} | B^- \rangle & \simeq & 
     \lambda_u^{(d)} \left[ 3 A_u(2,3/2) \right]
+ \lambda_c^{(d)} \left[3 A_c(2,3/2)\right] \ ,
 \\[0.3em]
 \langle \pi^+ \pi^- 
    | H_{\rm eff} | \bar B^0 \rangle & \simeq & 
     \lambda_u^{(d)}  \left[
     - A_u(0,1/2) 
     + A_u(2,3/2) 
\right] \ , 
\cr &&
 + \lambda_c^{(d)} \left[ - A_c(0,1/2) 
 + A_c(2,3/2)\right]
\\[0.3em]
\sqrt2 \,     \langle \pi^0 \pi^0 
    | H_{\rm eff} | \bar B^0 \rangle & \simeq & 
     \lambda_u^{(d)} 
     \left[A_u(0,1/2) 
     + 2 A_u(2,3/2)\right] 
\cr &&
 + \lambda_c^{(d)} \left[ A_c(0,1/2) + 2 A_c(2,3/2)\right] \ ,
\eeq
where the first argument denotes the total isospin $I$ of
the final state,  and the second argument denotes the
isospin $\Delta I$ of the operators in the weak effective
hamiltonian. For the charge-conjugated modes one has to 
replace $\lambda_i^{(d)}$ by $\lambda_i^{(d)*}$. 
Thus, the physical $B \to \pi\pi$ amplitudes fulfill
the well-known isospin relation
\beq
  A[B^- \to \pi^- \pi^0] - A[\bar B^0 \to \pi^0\pi^0]
    - A[\bar B^0 \to \pi^+\pi^-]/\sqrt2 &=&  0 \ .
\eeq
The latter is violated by small quark mass effects
and electromagnetic corrections which we discard in the following.

Notice that only electroweak penguin operators contribute 
to the amplitude $A_c(2,3/2)$. These operators  have very small Wilson
coefficients compared to those  in $A_u(2,3/2)$.
Actually, neglecting the tiny Wilson coefficients $C_7$ and $C_8$, one
can use Fierz identities to relate the matrix elements
of ${\cal O}_{9,10}$ with those of ${\cal O}_{1,2}$ and thus obtains
\beq
  \frac{A_c(2,3/2)}{A_u(2,3/2)} & \simeq & 
  \frac32 \, \frac{C_9 + C_{10}}{C_1+C_2} = {\cal O}(1\%) \ .
\label{FIERZ}
\eeq
The contribution from $A_c(2,3/2)$ can therefore be neglected 
in $B \to \pi\pi$ decays to a very good approximation.
This result is based on the structure of the 
effective electroweak hamiltonian and on the general isospin 
analysis only \cite{Fleischer:1995cg,Neubert:1998pt}.

Using this simplification one sometimes
introduces an intuitive parametrization, which refers
to the flavour topology (``penguin'', 
``tree'' or ``colour-suppressed'') of amplitudes that one would
obtain in the factorization approximation,
\beq
 \tilde P &=& \lambda_c^{(d)*} \, A_c(0,1/2)   \ , \cr
  \tilde T \, e^{i\gamma} 
    &=& \lambda_u^{(d)*} \left( A_u(0,1/2) - A_u(2,3/2)\right) \ , \cr
  \tilde C \, e^{i\gamma}
    &=& -\lambda_u^{(d)*} \left( A_u(0,1/2) + 2 A_u(2,3/2)\right) \ .
\label{topol}
\eeq
The ratios between these amplitudes can be parametrized in
terms of two moduli and two strong phases as
\beq
 d \, e^{i\theta} \equiv - \frac{\tilde P}{\tilde T} \, e^{i\gamma} 
 &=&
- \frac{1}{R_u} \,
  \frac{ A_c(0,1/2)}{A_u(2,3/2) - A_u(0,1/2)}
\eeq
as a measure for the ``penguin-to-tree ratio'', 
and
\beq
 x \, e^{i\Delta} \equiv \frac{\tilde C}{\tilde T} &=&
  \frac{A_u(0,1/2) + 2 \, A_u(2,3/2)}
       {A_u(2,3/2)- A_u(0,1/2)} 
\label{xparam}
\eeq
as a measure for the ratio of ``colour-suppressed'' to 
``colour-allowed'' tree amplitudes.
Beyond the factorization approximation
the notion of flavour topologies might be somewhat
misleading, whereas the classification in terms of
isospin amplitudes is more general.

\begin{figure}[tbph]
\begin{center}
  \psfig{file=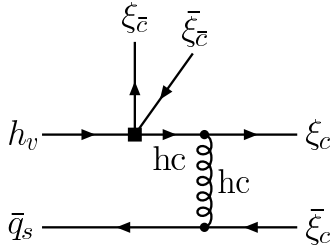, height=0.2\textwidth}
\end{center}
\caption{Example for a naively factorizing contribution to $B \to PP$.
 The labels indicate the scaling of the momentum modes.}
\label{fig:naive}
\end{figure}

The isospin amplitudes $A_i(I,\Delta I)$ contain contributions
from short-distance \mbox{dynamics}
(modes with large virtualities in the heavy quark limit) 
and long-distance dynamics (modes with virtualities
of order $\Lambda_{\rm QCD}^2$). The short-distance effects can be
treated in perturbative QCD, making use of the heavy-quark
expansion. The long-distance effects represent hadronic
uncertainties.
We therefore decompose every isospin amplitude as
\beq
  A_i(I,\Delta I) &=& A_i^{F}(I,\Delta I) + A_i^{NF}(I,\Delta I) \ .
\eeq
In  ``naive'' factorization, the amplitudes
$A_i^{F}$ can
be expressed in terms of electroweak Wilson coefficients and
hadronic decay constants and form factors. 
An example for a naively factorizing diagram is shown
in Fig.~\ref{fig:naive}, where we also indicate
the momentum scaling of external and internal lines
\cite{Beneke:2003pa}:
Here and in the following, ``s'' stands for soft momenta,
``c'' and ``$\bar{\rm c}$'' stand for collinear momenta
 (virtuality $~\Lambda^2$) in one or the other direction, 
  ``hc'' and $\overline{\rm hc}$'' for hard-collinear momenta
 (virtuality $~\Lambda m_b$) in one or the other direction, 
 and ``h'' for hard modes.

\begin{figure}[tbph]
\begin{center}
(a)  \psfig{file=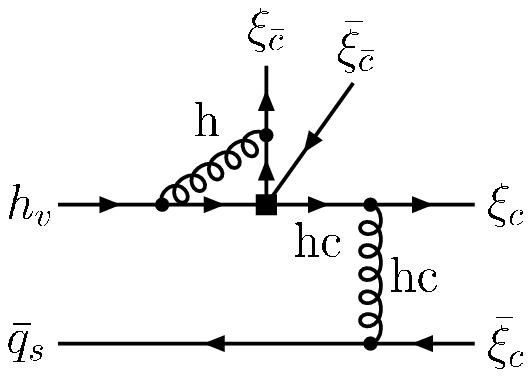, height=0.2\textwidth}
\hspace{3em}
(b)  \psfig{file=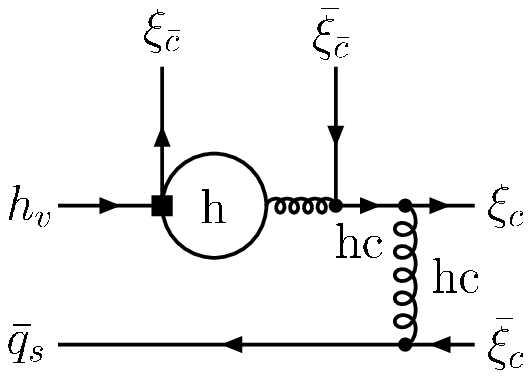, height=0.2\textwidth}
\end{center}
\caption{Examples for
(a) vertex and (b) penguin corrections to naive factorization
in $B \to PP$. The labels indicate the 
  scaling of the momentum modes.}
\label{fig:vertex}
\end{figure}

\begin{figure}[tbph]
\begin{center}
(a)  \psfig{file=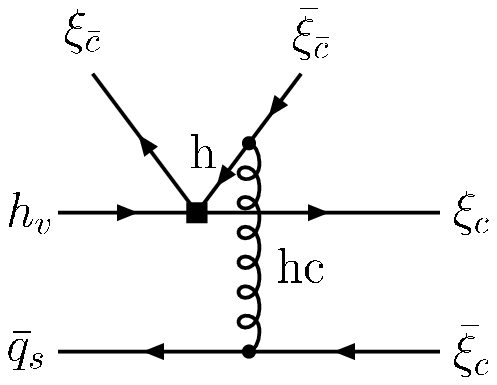, height=0.2\textwidth}
\hspace{3em}
(b)  
\psfig{file=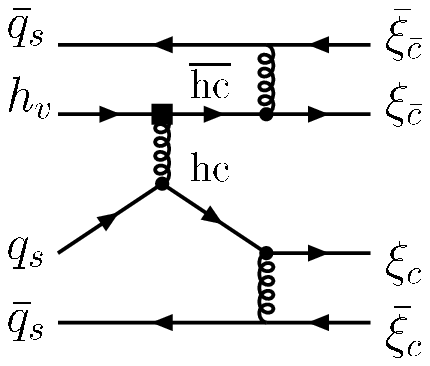, height=0.2\textwidth}
\end{center}
\caption{Hard-scattering contributions to $B \to PP$. 
  (a) Example for one-gluon exchange
  included in the
  BBNS analysis. (b) Example for (power-suppressed) higher-order diagram.
  The labels indicate the 
  scaling of the momentum modes.}
\label{fig:hardscatt}
\end{figure}

\begin{figure}[tbph]
\begin{center}
(a)  \psfig{file=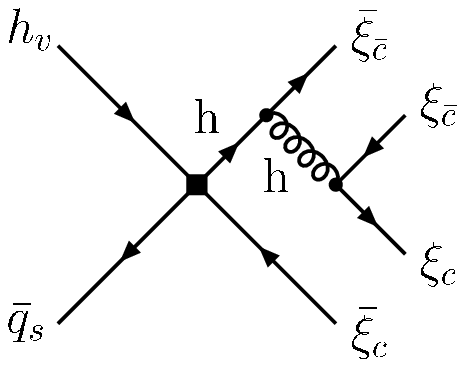, height=0.2\textwidth}
\hspace{3em}
(b)  \psfig{file=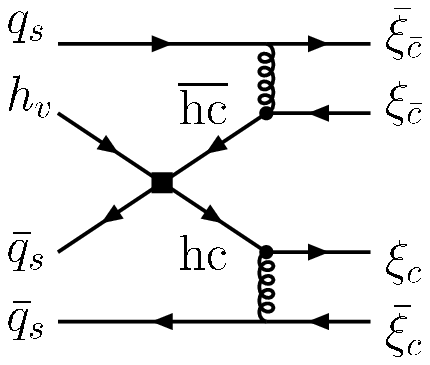, height=0.2\textwidth}
\end{center}
\caption{Annihilation contributions to $B \to PP$. 
  (a) Example for one-gluon exchange
  included in the
  BBNS analysis. (b) Example for higher Fock-state-contribution.}
\label{fig:anni}
\end{figure}

In the heavy-quark
limit, we can use QCD factorization (\ref{QCDF})
to improve the quantitative
description of the factorizable part $A_i^{F}$. 
The contributions to the first term in  (\ref{QCDF})
come from vertex corrections and penguin contractions
as shown in Fig.~\ref{fig:vertex}, and from hard spectator scattering
as shown in Fig.~\ref{fig:hardscatt}(a).
Non-factorizable effects arise from power corrections
in the $\Lambda_{\rm QCD}/m_b$ expansion.
They include the annihilation topologies (see Fig.~\ref{fig:anni}),
and cannot be calculated in a reliable way at present.

\subsection{Factorizable contributions}

In \cite{Beneke:1999br} the contributions related to
$T_{\rm I}$ and  $T_{\rm II}$ in (\ref{QCDF}) are
expressed in terms of parameters $a_i =a_{i,\rm I}+a_{i,\rm II}$
($a_{i,\rm II}$ will be restricted to the heavy-quark limit,
whereas in $a_{i,\rm I}$ the terms proportional
to $r_\chi^\pi =m_\pi^2/m_q m_b$ are kept).
The factorizable amplitudes $A_i^F$ 
read
\beq
  A_u^F(0,1/2) &=& \frac{A_{\pi\pi}}{6} \left(
    4 a_{1} - 2a_{2}
    + 6 a_{4}^u 
    + 3 a_{7} - 3 a_{9} + 3 a_{10}^u
   + r_\chi^\pi \,(6 a_{6}^u + 3 a_{8}^u) 
   \right)
\cr &=& \left( 0.626^{+0.027}_{-0.022} + i \, 0.007^{+0.018}_{-0.010} \right) 
        \, A_{\pi\pi} \
\cr &&  - \ 0.037  
    \left[\frac{0.28}{F_0^{B\to\pi}} 
          \, \frac{350~{\rm MeV}}{\lambda_B} 
          \left(
          \frac{\langle \bar u^{-1}\rangle_\pi}{3.3}
         \right)^2 \right] 
\, A_{\pi\pi} \ ,
\nonumber \\[0.2em]
   A_u^F(2,3/2) &=& -\frac{A_{\pi\pi}}{6}
\left( 2a_{1} + 2a_{2}
    - 3 a_{7} + 3 a_{9} + 3 a_{10}^u
   + 3 r_\chi^\pi \,a_{8}^u 
   \right)
\cr &=& - \left( 0.350^{+0.007}_{-0.008} - i \, 0.021^{+0.003}_{-0.004} \right) \, A_{\pi\pi} \ \cr &&
- \ 0.023 
 \left[\frac{0.28}{F_0^{B\to\pi}}
          \, \frac{350~{\rm MeV}}{\lambda_B}
          \left(
          \frac{\langle \bar u^{-1}\rangle_\pi}{3.3}
         \right)^2 \right]  
  \, A_{\pi\pi}
\nonumber \ , \\[0.2em]
  A_c^F(0,1/2) &=& \frac{A_{\pi\pi}}{2} \left(
    2 a_{4}^c 
    + a_{7} -  a_{9} +  a_{10}^c
   + r_\chi^\pi \,(2 a_{6}^c +  a_{8}^c) 
   \right) 
\cr &=& 
- \left( 0.086^{+0.004}_{-0.004} + i \, 0.013^{+0.001}_{-0.000} \right) \, A_{\pi\pi}\ \cr &&
+ \ 0.001 
 \left[\frac{0.28}{F_0^{B\to\pi}}
          \, \frac{350~{\rm MeV}}{\lambda_B}
          \left(
          \frac{\langle \bar u^{-1}\rangle_\pi}{3.3}
         \right)^2 \right]   \, A_{\pi\pi}
\nonumber \ , \\[0.2em]
   A_c^F(2,3/2) &=& -\frac{A_{\pi\pi}}{2}
\left( 
    - a_{7} + a_{9} + a_{10}^c
   + r_\chi^\pi \,a_{8}^c 
   \right) \simeq  0.004 \, A_{\pi\pi} \ ,
\label{AF}
\eeq
where we introduced
$$
 A_{\pi\pi} = \frac{i G_F}{\sqrt2} \, (m_B^2-m_\pi^2) \,
 F_0^{B \to \pi}(m_\pi^2) \, f_\pi \ ,
$$
which determines the overall normalization of the factorizable
amplitudes, and $F_0^{B \to \pi}(q^2)$ is the scalar $B \to \pi$
transition form factor.

The error in the first term of (\ref{AF})
refers to the variation of the factorization
scale $\mu$ between $m_b/2$ and $2 m_b$ in the vertex and penguin
graphs. The second term denotes the central value
for the hard-scattering contribution which has a large uncertainty
related to the first inverse moment $\lambda_B^{-1}$
of the light-cone distribution amplitude of the $B$\/ meson.
(There are more sources of parametric uncertainties,
in particular the scale-dependence of the hard-scattering
term, see the numerical discussion in \cite{Beneke:1999br}.
Notice that for the hard-scattering terms,
we considered the electroweak
Wilson coefficients at the scale $\mu=m_b$.)

In the following numerical discussion we will refer to
the central values of the $A_i^F$ only. Of course, a
significant change in the numerical values of  $A_i^F$
would also influence our conclusions about the size of
$A_i^{NF}$. In particular, one has to keep in mind that
quantities like the ``colour-suppressed'' amplitude $C$
defined in (\ref{topol}) involve large cancellations in
naive factorization, and therefore are particularly
sensitive to the precise value of hard-scattering 
and non-factorizable corrections.
The question  whether the variation of all possible
input parameters in the BBNS approach 
within ``reasonable'' ranges could reproduce the experimental
data has already been studied in \cite{Charles:2004jd}.\footnote{For 
instance, using the ``large-$a_2$'' scenario
in \cite{Beneke:2003zv}, where $\langle \bar u^{-1}\rangle_\pi=4.2$,
$\lambda_B = 200$ MeV, and $F_0^{B \to \pi}=0.25$, the 
hard-scattering correction in $a_{i,\rm II}$ increase by
a factor of 3.} Here we take the point of view that
the default values in \cite{Beneke:1999br} give a reliable
prediction for the {\em factorizable}\/ contributions, whereas
the BBNS analysis of non-factorizable terms  (including the
variation of hadronic input parameters) is considered as
one option among different alternatives.

The power-corrections to
the hard-scattering parameters $a_{i,\rm II}$,
and the annihilation parameters
$b_i$ are considered
as part of the unknown functions $A_i^{NF}$, which we parametrize
as
\beq
  A_i^{NF}(I,\Delta I) &:=& r_i(I,\Delta I) \, 
   e^{i \phi_i(I,\Delta I)} \,  A_{\pi\pi}
\label{eq:general}
\eeq
with $r_i > 0$ and an arbitrary phase $\phi_i$.
(As explained above, one can safely set $A_c^{NF}(2,3/2)=0$:
Even if we allow for an order-of-magnitude enhancement 
with respect to its factorizable counterpart in (\ref{AF}),
we would only get an  ${\cal O}(1\%)$ correction to $A_u(2,3/2)$.)

\subsection{Non-factorizable effects from $B \to \pi\pi$ data}

Our general parametrization of non-factorizable effects 
introduces seven adjustable para\-me\-ters $r_u(0,1/2)$,
$\phi_u(0,1/2)$,  $r_u(2,3/2)$,
$\phi_u(2,3/2)$,  $r_c(0,1/2)$,
$\phi_c(0,1/2)$, and $F_0^{B \to \pi}$.
On the other hand, if we neglect the
tiny contribution from $A_c(2,3/2)$, on general grounds, we have only
five relevant parameters to describe three  complex isospin
amplitudes for $B \to \pi\pi$ (one overall phase is
not observable). Consequently, our parametrization contains
some redundancy, which we will keep for the moment.
Later we will consider different constrained scenarios, where
the number of parameters is less than 5.

\begin{table}[hbt]
\begin{center}
  \begin{tabular}{|l||c | c | c || c|}
\hline
      Observable & BaBar & Belle & CLEO & Average \\
\hline \hline
  ${\cal B}[B^0 \to \pi^+\pi^-]$ & 
  $4.7 \pm 0.6 \pm 0.2$ &
  $4.4 \pm 0.6 \pm 0.3$ &
  $4.5 ^{+1.4 + 0.5}_{-1.2-0.4}$ &
  $4.55 \pm 0.44$
\\ 
  ${\cal B}[B^+ \to \pi^+\pi^0]$ & 
  $5.5^{+1.0}_{-0.9} \pm 0.6   $ &
  $5.0 \pm 1.2 \pm 0.5         $ &
  $4.6 ^{+1.8 + 0.6}_{-1.6-0.7}$ &
  $5.18 ^{+0.77}_{-0.76}$
\\ 
  ${\cal B}[B^0 \to \pi^0\pi^0]$ & 
  $2.1 \pm 0.6 \pm 0.3         $ &
  $1.7 \pm 0.6 \pm 0.2         $ &
  $< 4.4 $ &
  $1.90 \pm 0.47 $ 
\\
\hline
 $C_{\pi\pi}^{+-}$ &
 $-0.19 \pm 0.19 \pm 0.05$ &
 $-0.58 \pm 0.15 \pm 0.07$ &
 -- &
 $-0.46 \pm 0.13$
\\
 $S_{\pi\pi}^{+-}$ &
 $-0.40 \pm 0.22 \pm 0.03$ &
 $-1.00 \pm 0.21 \pm 0.07$ &
 -- &
 $-0.73 \pm 0.16$
\\
\hline
 ${\cal A}_{\rm CP}[\pi^+\pi^0]$ &
 $-0.03 ^{+0.18}_{-0.17} \pm 0.02$ &
 $-0.14 \pm 0.24 ^{+0.05}_{-0.04}$ &
 -- &
 $-0.07 \pm 0.14$
\\
\hline 
  \end{tabular}
\end{center}
\caption{Experimental results on $B \to \pi\pi$ observables 
\cite{Charles:2004jd}.
All branching ratios are CP\/-averaged and
quoted in units of $10^{-6}$. The sign convention for CP\/-asymmetries
is defined in the text.}
\label{tab:exp}
\end{table}

Our strategy to infer information on the non-factorizable 
parameters from experimental data is to produce 
sets of random parameter values, and calculate the 
corresponding $\chi^2$\/-value by comparing the theoretical
branching ratios and CP asymmetries with the experimental
measurements in Table~\ref{tab:exp}~\footnote{ 
Updated results for charmless non-leptonic $B$\/-decays
have been presented by the BaBar and Belle collaborations in 
\cite{Abe:2004mp,Chao:2004jy,Aubert:2004kw,Aubert:2004kn}. 
The new data are consistent with the numbers
presented in \cite{Charles:2004jd}. We have checked that
they would only lead to marginal effects in our numerical
analysis, and our qualitative conclusions remain unchanged. 
With the new data 
the compatibility with (\ref{constr}), see below,
appears to be slightly improved.}.
We follow the sign convention of \cite{Charles:2004jd},
\beq
  A_{\rm CP}[\pi^+\pi^0] &=& 
 \frac
{\Gamma[\bar B^- \to \pi^-\pi^0]-\Gamma[\bar B^+ \to \pi^+\pi^0]}
{\Gamma[\bar B^- \to \pi^-\pi^0]+\Gamma[\bar B^+ \to \pi^+\pi^0]}
\ ,
\eeq
and
\beq
&& S_{\pi\pi}^{+-} =  
\frac
{2 {\rm Im} \lambda_{\pi\pi}}
{1 + |\lambda_{\pi\pi}|^2}
 \ , \qquad
C_{\pi\pi}^{+-} =  
\frac
{1 - |\lambda_{\pi\pi}|^2}
{1 + |\lambda_{\pi\pi}|^2}
 \ , 
\\[0.2em]
&& \lambda_{\pi\pi} = \frac{q}{p} \, 
\frac{{\cal A}[\bar B^0 \to \pi^+\pi^-]}{{\cal A}[B^0 \to \pi^+\pi^-]}
\simeq
e^{-2i\beta}  \, 
\frac{{\cal A}[\bar B^0 \to \pi^+\pi^-]}{{\cal A}[B^0 \to \pi^+\pi^-]}
\eeq
The so-obtained $\chi^2$ distributions enable us to
investigate the generic size and importance of non-factorizable parameters.
To generate the sample, 
we assume uniform distributions of parameter values in the
following ranges
\beq
  0.23 \leq F_0^{B \to \pi}(m_\pi^2) \leq 0.33 \ , \nonumber \\[0.2em]
  0 \leq r_{u,c}(I,\Delta I) \leq 1.0 \ , \nonumber \\[0.2em]
  0^\circ \leq \phi_{u,c}(I,\Delta I) \leq 360^\circ \ ,
\eeq
The bound on the scalar form factor is the main theoretical bias.
We used a rather conservative estimate of the theoretical 
uncertainties, which contains the central value $0.28$ used
in \cite{Beneke:1999br} as well as
a recent update $0.26$ for this quantity in the framework 
of QCD sum rules \cite{Ball:2004ye}.
The upper bound on $r_{u,c}(I,\Delta I)$ will turn out to be
sufficiently large not to induce an additional bias.

\begin{figure}[tbph]
\small
\begin{tabular}{cc}
\multicolumn{2}{c}{\small $F_0^{B \to \pi}$}\\[-0.3em]
\multicolumn{2}{c}{
  \psfig{file=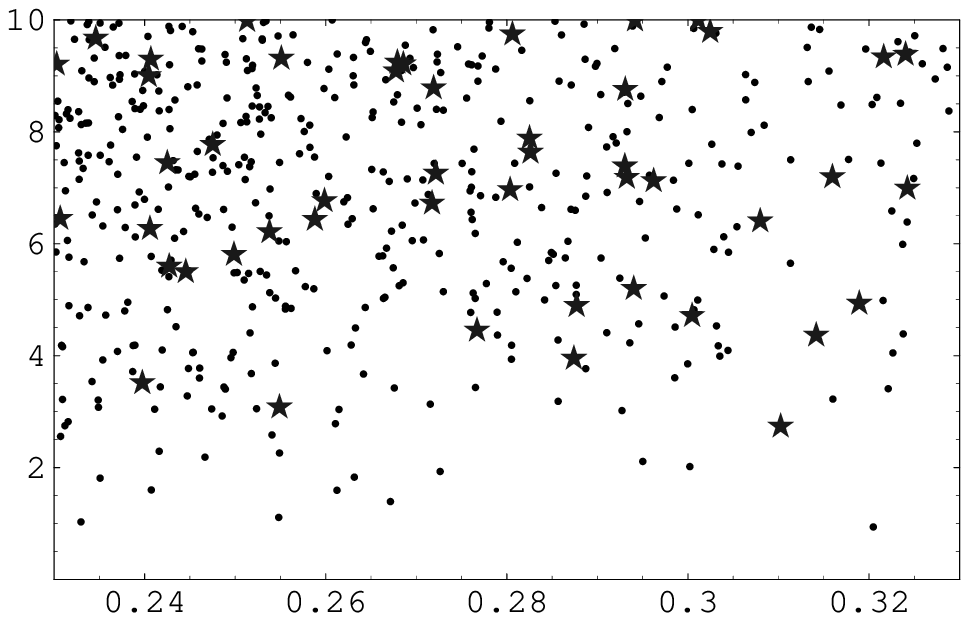, width=0.46\textwidth}}
\\[0.1em]
\small $r_u(0,1/2)$ & \small $\phi_u(0,1/2)$ \\[-0.3em]
\psfig{file=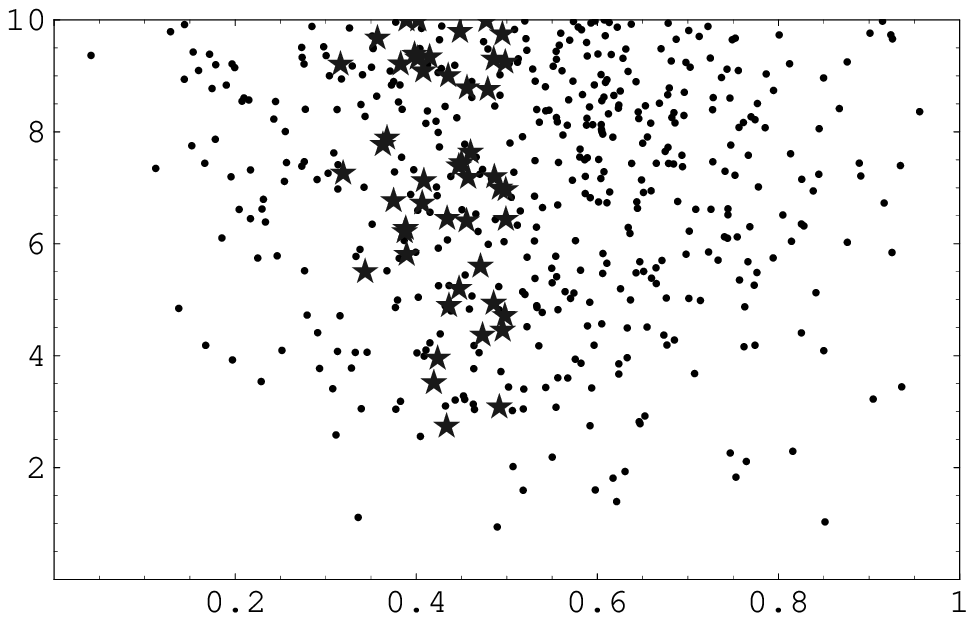, width=0.46\textwidth}
&
\psfig{file=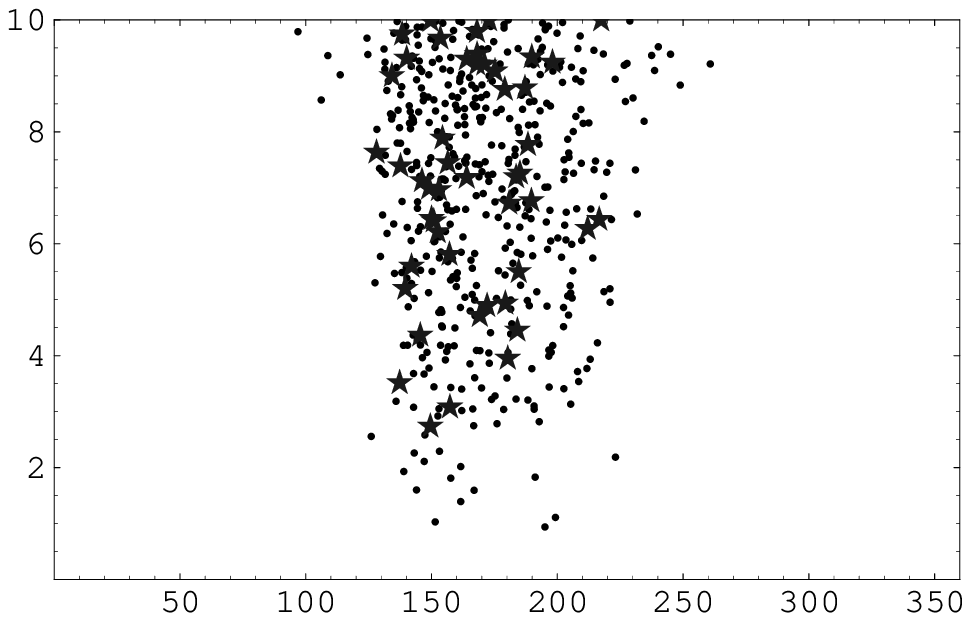, width=0.46\textwidth}
\\[0.1em]
\small $r_c(0,1/2)$ & $\phi_c(0,1/2)$ \\[-0.3em]
\psfig{file=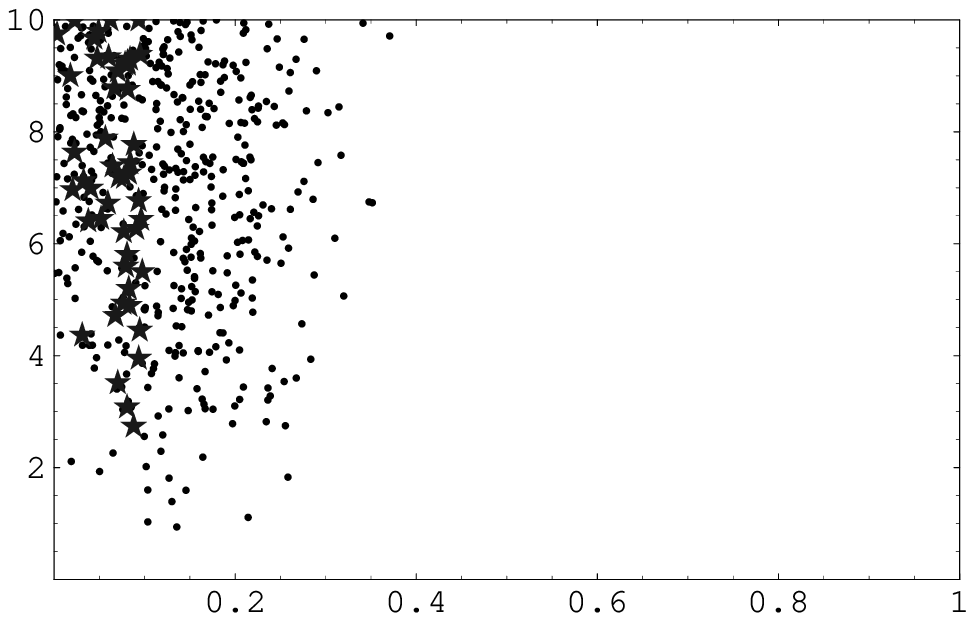, width=0.46\textwidth}
&
\psfig{file=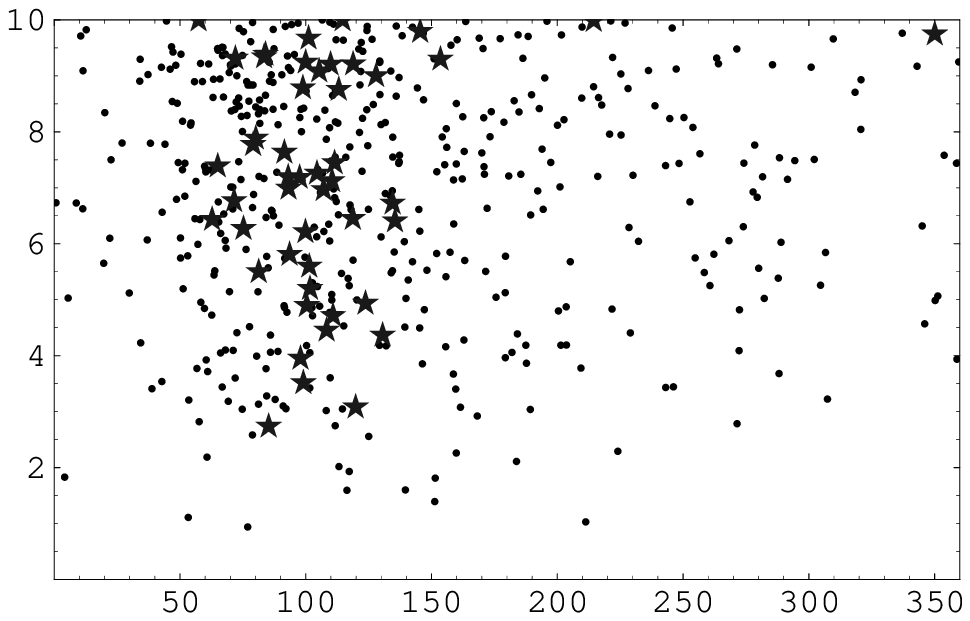, width=0.46\textwidth}
\\[0.1em]
$r_u(2,3/2)$ & $\phi_u(2,3/2)$ \\[-0.3em]
\psfig{file=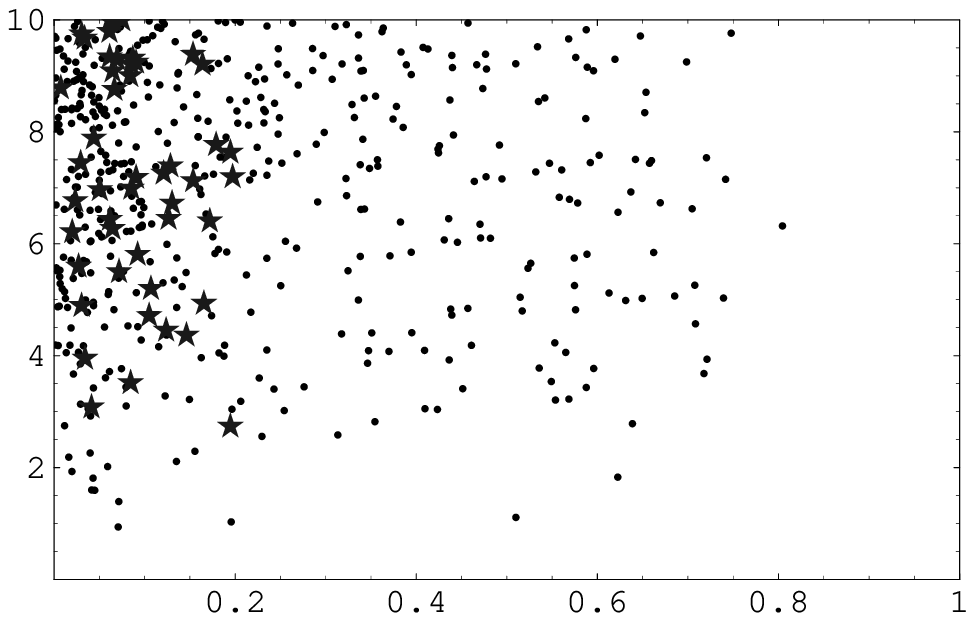, width=0.46\textwidth}
&
\psfig{file=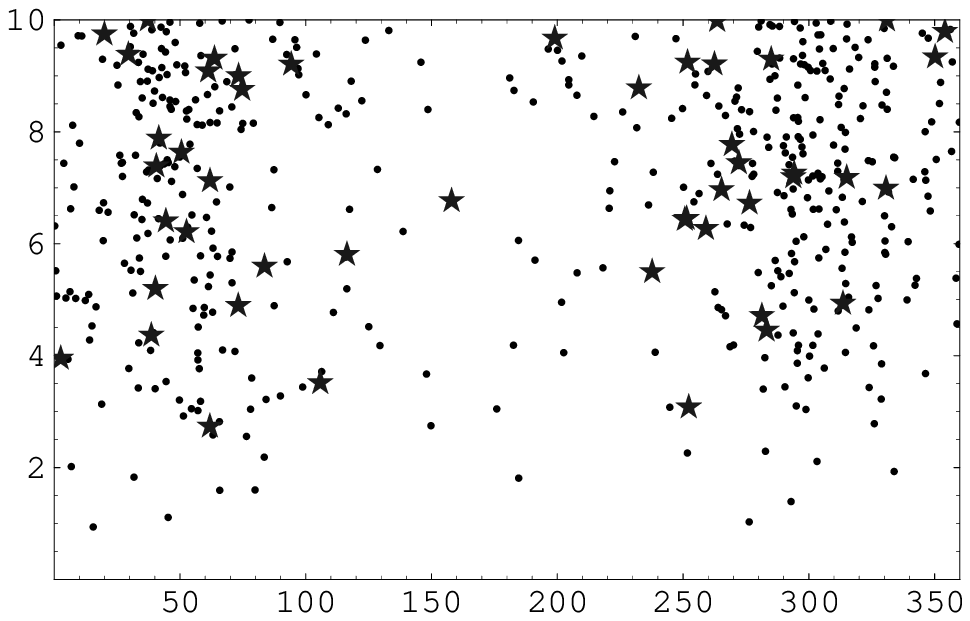, width=0.46\textwidth}
\end{tabular}
\normalsize
\caption{
Comparison of $\chi^2$ values for a random
sample of non-factorizable parameters:
Dots for unconstrained scenario, stars
for combinations that fulfill (\ref{constr}).
}
\label{fig:par}
\end{figure}

Already for the unconstrained scenario we find some interesting
patterns, see Fig.~\ref{fig:par}, where
we have plotted the $\chi^2$ value against
each of the free parameters for a sample of 500 points 
with $\chi^2 < 10$:
\begin{itemize}
  \item The value of the scalar $B \to \pi$ form factor
        is not constrained by the data\footnote{
        It would therefore be interesting to independently measure the value of
        the form factor from $B \to \pi\ell\nu$ decay
        for given value of $|V_{ub}|$.}

  \item The distribution of the parameter $r_u(0,1/2)$ is rather
        broad, with a slight preference for $r_u(0,1/2)=0.5\pm0.5$.
        The corresponding phase take values in the range 
        $120^\circ< \phi_u(0,1/2)<240^\circ$. (Notice that for values
        of $\phi_u(0,1/2)$ around $180^\circ$ the non-factorizable
        contributions reduce the value of the
        ``colour-allowed'' tree amplitude, which accomodates
        the experimental fact that the parameter $x$ in 
        (\ref{xparam}) is large.)
  
  \item We further find $r_c(0,1/2)<0.35$, and $r_u(2,3/2)<0.8$.
        The corresponding phases $\phi_c(0,1/2)$ 
        and $\phi_u(2,1/2)$ show no pronounced preference.
\end{itemize}

Of course, there are correlations between the different parameters.
For instance, there is no solution with small $\chi^2$, where
{\it all}\/ parameters $r_i(I,\Delta I)$ are small (relative
to the factorizable terms), in other words the {\em default scenario}\/
in the BBNS approach is disfavoured by the data (in accordance with
similar conclusions in \cite{Buras:2004ub,Charles:2004jd,Ali:2004hb}).
For illustration we marked solutions which 
simultaneously fulfill
\beq
  r_u(0,1/2) < 0.5 \ ,  \qquad 
  r_u(2,3/2) < 0.2 \ ,  \qquad \mbox{and} \qquad
  r_c(0,1/2) < 0.1 \ ,
\label{constr}
\eeq
and which, in view of the large parametric uncertainties 
related to hard-scattering and annihilation contributions
(see below) may still be viewed as 
more or less compatible with the BBNS approach (in the 
spirit of \cite{Charles:2004jd}). This still yields values
of $\chi^2 \simeq 3$.

\subsection{Non-factorizable effects from ``BBNS''}

In the BBNS approach non-factorizable effects
arise through chirally enhanced power-corrections
which are identified from endpoint-divergent
convolution integrals, appearing in the diagrammatic approach.

\subsubsection{Hard-scattering contributions}

One source of non-factorizable power-corrections
are so-called hard spectator-scattering
diagrams, where a gluon connects the spectator quark to the
short-distance decay process.
Fig.~\ref{fig:hardscatt}(a) (together with an analogous diagram
where the gluon is attached to the other collinear quark)
has been considered in \cite{Beneke:1999br}. 
On the other hand, Fig.~\ref{fig:hardscatt}(b) represents
an example of a higher-order diagram (which is not included in the
BBNS approach), which involves a multi-particle Fock state in
the $B$\/ meson and which is power-suppressed (see below).

Apart from a factorizable part that determines  the
heavy quark limit, explicit calculation shows that
the diagram in Fig.~\ref{fig:hardscatt}(a) 
gives rise to chirally-enhanced power-suppressed
endpoint divergences.
At the considered order in the diagrammatic expansion,
these endpoint divergences enter through the quantity
\beq
  X_H^\pi
 &\equiv & \int_0^{1-\Lambda_h/m_B} \, \frac{u}{1-u} \, \phi_p^\pi(u)
\label{XHdef} \ ,
\eeq
where $\phi_p^\pi(u)\simeq 1$ 
is a twist-3 light-cone distribution amplitude
of the pion. We regularized the integral by means of
an ad-hoc cut-off, because the integral does not
converge for $u \to 1$. Since the twist-3 distribution amplitudes
are normalized to a term proportional to the quark condensate
(i.e.\ the ratio of Goldstone boson and quark masses, $m_M^2/m_q$),
their contributions are numerically large (``chirally enhanced'').
To study the phenomenological impact of these terms,
the authors of \cite{Beneke:1999br} propose to parametrize the quantity on the left-hand
side in terms of 
\beq
  X_H^\pi 
  &=& (1 + \rho_H \, e^{i \varphi_H}) \, \ln \frac{m_B}{\Lambda_H} 
\ , \qquad \rho_H \sim {\cal O}(1)
\eeq
with $\varphi_H$ being an arbitrary phase,
whereas $\Lambda_H \approx 0.5$~GeV. 

Inserting the default values for the hadronic parameters
in the BBNS analysis, we obtain
\beq
  A_u^{\rm NF}[0,1/2] |_{\rm hs} &\simeq &
   -0.030 \, A_{\pi\pi} \, (1 + \rho_H \, e^{i \varphi_H})
 \left[\frac{0.28}{F_0^{B\to\pi}}
          \, \frac{350~{\rm MeV}}{\lambda_B}
         \,\frac{\langle \bar u^{-1}\rangle_\pi}{3.3}
         \right]  
\ ,\\[0.2em]   
  A_u^{\rm NF}[2,3/2] |_{\rm hs} &\simeq &
   -0.018 \, A_{\pi\pi} \, (1 +  \rho_H \, e^{i \varphi_H})
 \left[\frac{0.28}{F_0^{B\to\pi}}
          \, \frac{350~{\rm MeV}}{\lambda_B}
         \,\frac{\langle \bar u^{-1}\rangle_\pi}{3.3}
         \right]    
\ ,\\[0.2em]   
  A_c^{\rm NF}[0,1/2] |_{\rm hs} &\simeq &
   0.001 \, A_{\pi\pi} \, (1 +  \rho_H \, e^{i \varphi_H})
 \left[\frac{0.28}{F_0^{B\to\pi}}
          \, \frac{350~{\rm MeV}}{\lambda_B}
         \,\frac{\langle \bar u^{-1}\rangle_\pi}{3.3}
         \right]    
\ ,\\[0.2em]   
  A_c^{\rm NF}[2,3/2] |_{\rm hs} &\simeq & 0 \ .
\eeq
Comparing with the general parametrization (\ref{eq:general}), we
deduce that in the BBNS approach
$r_u(I,\Delta I)$ are expected to receive
contributions of the order 5\% (up to 10\% for the
``large $a_2$ scenario'') from hard spectator-scattering, 
whereas the corresponding effect for $r_c(I,\Delta I)$
seems to be negligible.

In any case the above procedure is understood to
only give a rough idea about the typical size of non-factorizable
contributions for {\it individual}\/ decay amplitudes.
The origin and size of strong re-scattering 
phases remains unclear. In particular, treating $\rho_H$ and
$\varphi_H$ as universal parameters, one induces model-dependent
{\it correlations}\/ between
non-factorizable effects in different isospin amplitudes.
Substantially different strong interaction phases for, say, 
$I=0$ and $I=2$ final states can arise from 
more complicated diagrams with additional
quark lines (i.e.\ higher Fock states). In the diagrammatic approach 
along the lines of BBNS, these effects can only show up
at higher orders in the diagrammatic expansion.
Notice that in the case of non-factorizable endpoint configurations, 
``higher-order'' diagrams are not necessarily suppressed by powers
of the strong coupling constant.
The universality of strong phases arising from hard-spectator
scattering is therefore {\it not}\/ a generic feature of 
QCD factorization, but represents an {\em additional}\/ 
phenomenological assumption in the BBNS approach. 
The analysis \cite{Charles:2004jd} uses universal BBNS parameters for
{\it all}\/ isospin amplitudes. Different phases in
different isospin amplitudes can still be
generated by combining the non-factorizing pieces from hard
scattering ($X_H$) and from annihilation ($X_A$ see below) which are treated
as independent complex parameters.
Nevertheless, a model-dependent correlation between non-factorizable 
contributions to different $B \to \pi\pi$ and $B \to \pi K$
amplitudes remains. The implications of that 
analysis should therefore be interpreted with some care.

\subsubsection{Scenario 1: Dominance of hard spectator scattering}

If we assume that non-factorizable effects from power-suppressed contributions
to hard spectator scattering (or alternatively, significant changes of
hadronic input parameters in the factorizable pieces)
are the main source for $A^{\rm NF}(I,\Delta I)$,
we may define a phenomenological approximation, 
where we put $r_c(0,1/2)=0$. In addition, we might
either fix the form factor value, $F_0 = 0.26$ (Scenario 1a), 
or assume that the phases
$\varphi_H$ are universal 
such that $\phi_u(0,1/2)=\phi_u(2,1/2)$ (Scenario 1b).

\begin{figure}[tbph]
\small
\begin{tabular}{cc}
\small $r_u(0,1/2)$ & \small $\phi_u(0,1/2)$ \\[-0.3em]
\psfig{file=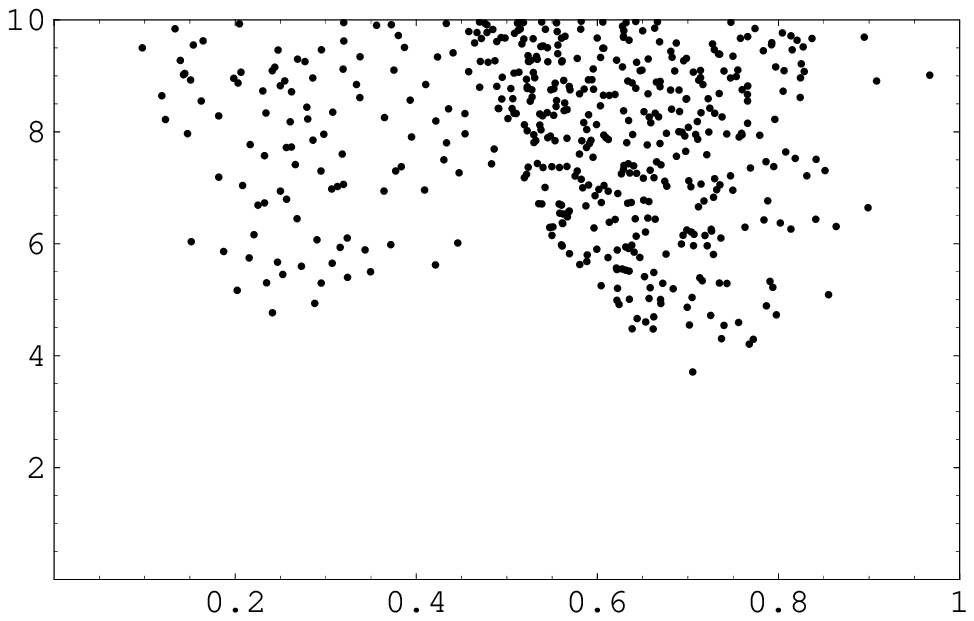, width=0.46\textwidth}
&
\psfig{file=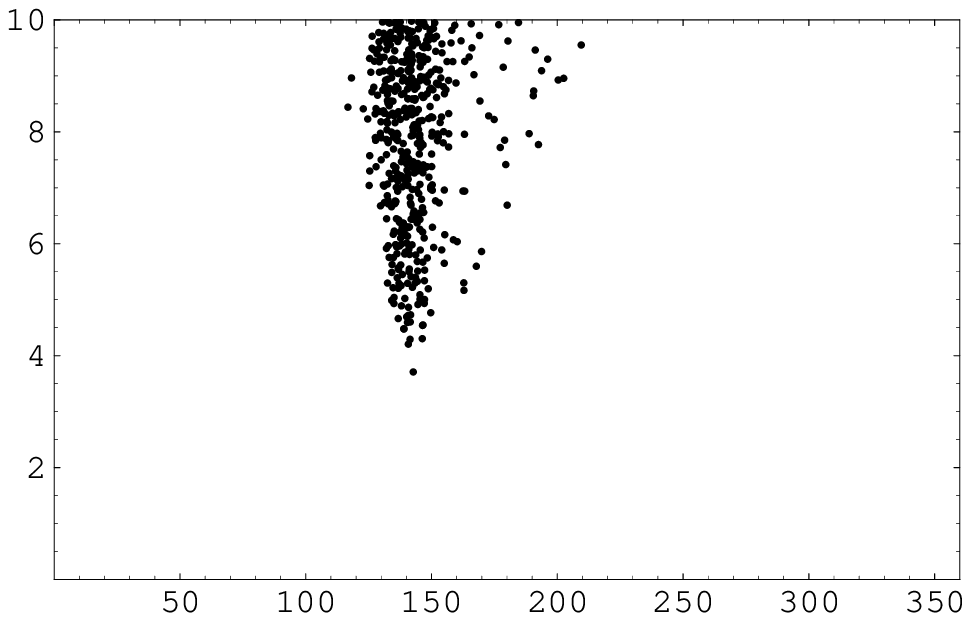, width=0.46\textwidth}
\\[0.1em]
\small $r_u(2,3/2)$ & $\phi_u(2,3/2)$ \\[-0.3em]
\psfig{file=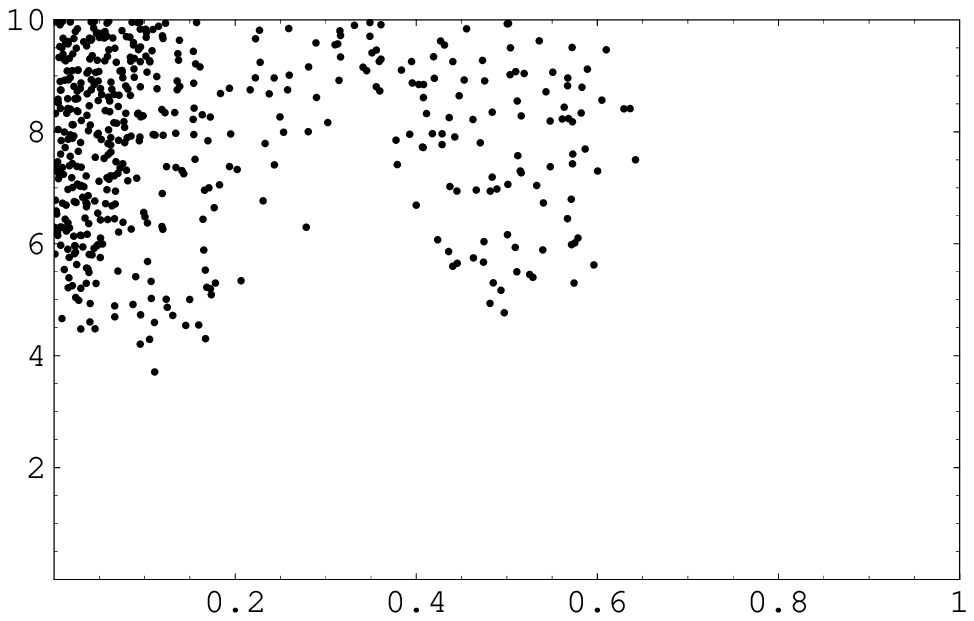, width=0.46\textwidth}
&
\psfig{file=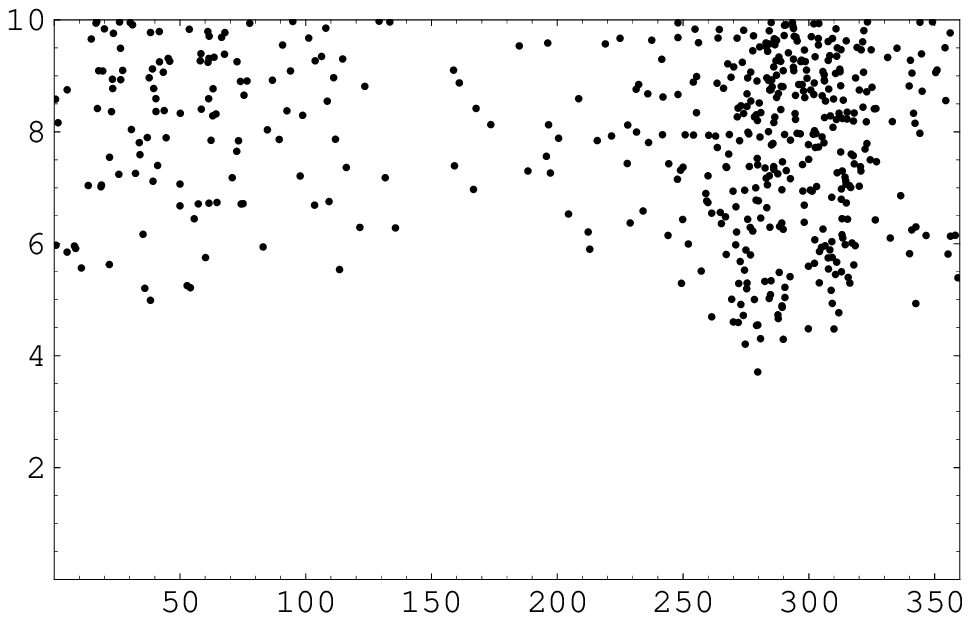, width=0.46\textwidth}
\end{tabular}
\normalsize
\caption{Comparison of $\chi^2$ values for a random
sample of non-factorizable parameter combinations, using
$r_c(0,1/2)=0$ and $F_0^{B \to \pi}=0.26$ (Scenario 1a).}
\label{fig:par_scenario1a}
\end{figure}

\begin{figure}[tbph]
\small
\begin{tabular}{cc}
\small $r_u(0,1/2)$ & \small $\phi_u(0,1/2)$ \\[-0.3em]
\psfig{file=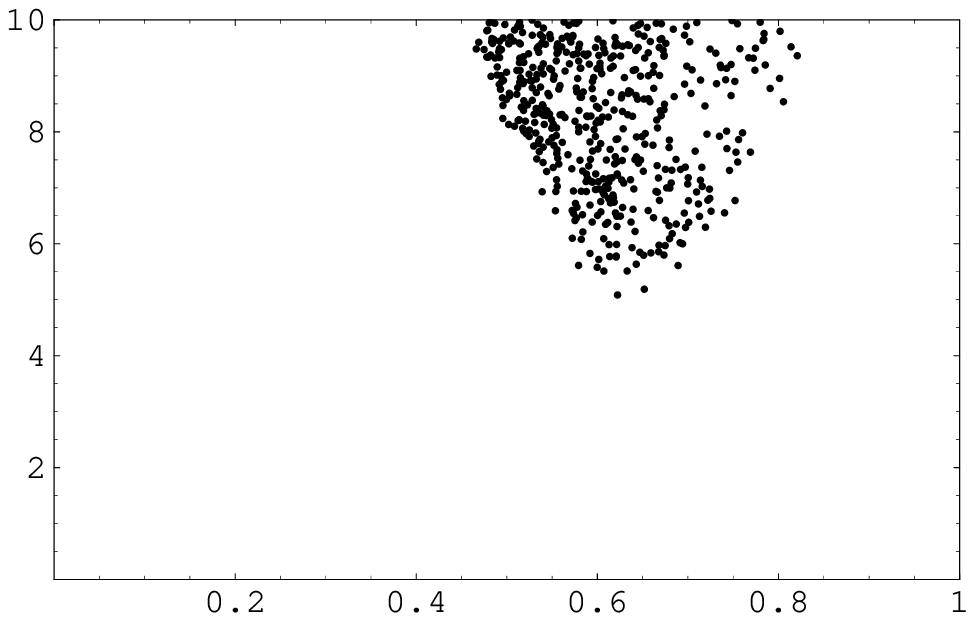, width=0.46\textwidth}
&
\psfig{file=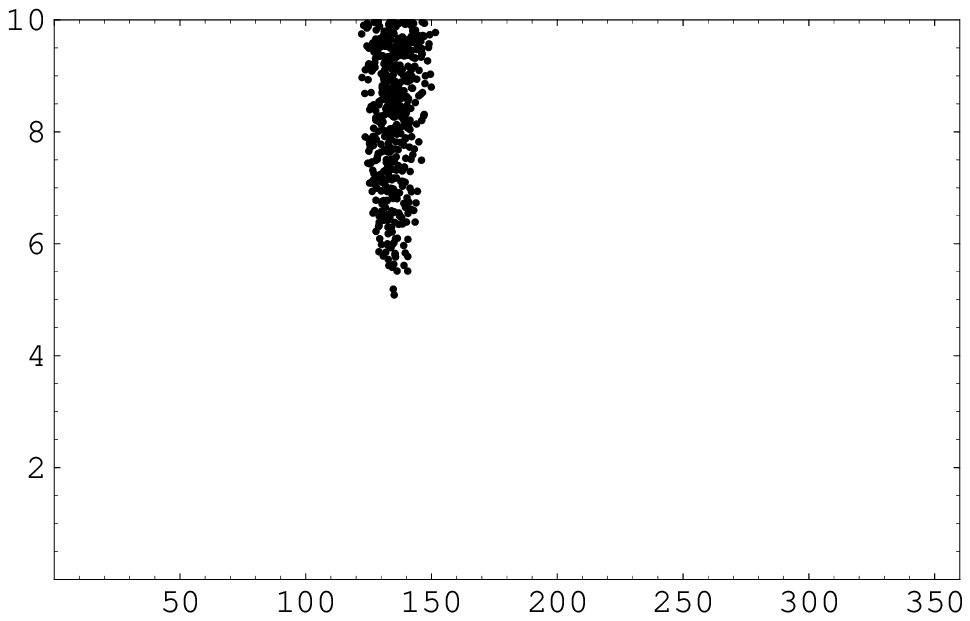, width=0.46\textwidth}
\\[0.1em]
\small $r_u(2,3/2)$ & $F_0^{B \to \pi}$ \\[-0.3em]
\psfig{file=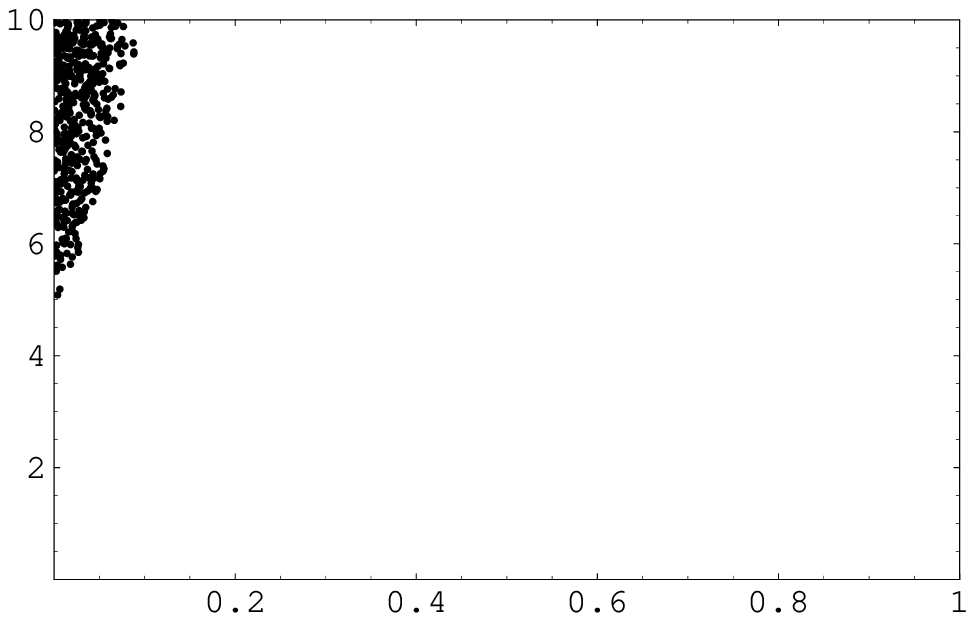, width=0.46\textwidth}
&
\psfig{file=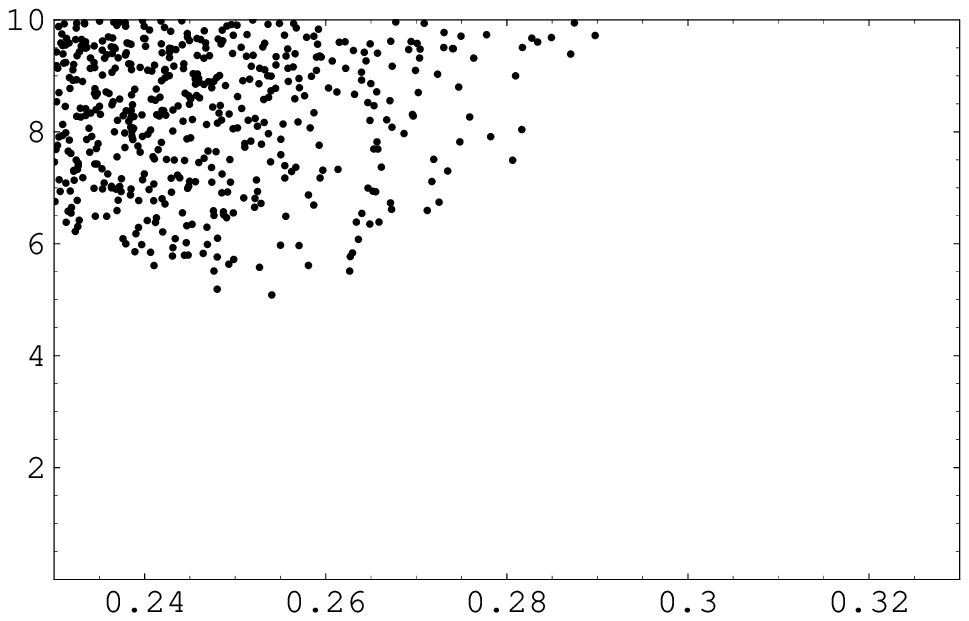, width=0.46\textwidth}
\end{tabular}
\normalsize
\caption{Comparison of $\chi^2$ values for a random
sample of non-factorizable parameter combinations, using
$r_c(0,1/2)=0$ and $\phi_u(2,3/2)=\phi_u(0,1/2)$ (Scenario 1b).}
\label{fig:par_scenario1b}
\end{figure}

Repeating the analysis
of the $B \to \pi\pi$ data with these additional constraints,
we obtain the situation illustrated in  
Figs.~\ref{fig:par_scenario1a} and \ref{fig:par_scenario1b}.
 Scenario~1a still gives a more or less reasonable description
with $\chi^2 \geq 4$ for six experimental observables and
four adjustable parameters. The situation 
in Scenario~1b is similar, but somewhat worse.
In both cases we need rather large non-factorizable amplitudes
with either $r_u(0,1/2) > 0.5$ or $r_u(2,3/2) > 0.3$.
$\phi_u(0,1/2)$ is rather constrained in both cases.

\subsubsection{Annihilation topologies}

Another source of non-factorizable,
power-suppressed contributions
to $B \to PP$ are annihilation 
topologies, see Fig.~\ref{fig:anni}.
They receive the same chiral enhancement as the non-factorizable
hard-scattering pieces.

In the BBNS approach, the flavour-changing sub-process
for annihilation topologies is $bq \to q\bar q$,
and consequently, it can only contribute to $I =0$
(for $B \to \pi\pi$) or $I=1/2$ (for $B \to \pi K$)
decay amplitudes (again this statement is only true as long as we
do not consider higher Fock states).
The order of magnitude for annihilation effects is estimated
in a similar way as for the hard-scattering terms in the
previous section, introducing the quantity
\beq
  X_A^\pi 
  &=& (1 + \rho_A \, e^{i \varphi_A}) \, \ln \frac{m_B}{\Lambda_H} 
\ , \qquad \rho_A \sim {\cal O}(1) \ .
\label{XA}
\eeq
The corresponding contributions to the non-factorizable 
$B \to \pi\pi$ amplitudes read
\beq
  A_u(0,1/2) |_{\rm ann} 
 &\simeq& 0.01 \,
 (1 + 1.0 \rho_A  e^{i\varphi_A} -  0.7 \rho_A^2 e^{2 i\varphi_A}) 
\, A_{\pi\pi}
\ ,\\[0.2em]
  A_c(0,1/2) |_{\rm ann} 
 &\simeq& -0.01 \, 
(1 + 2.0 \rho_A e^{i\varphi_A} + 0.9 \rho_A^2 e^{2i\varphi_A} ) 
\,  A_{\pi\pi} \ .
\eeq
Comparing with the general parametrization (\ref{eq:general}), we
deduce that in the BBNS approach
$r_{u,c}[0,1/2]$ may receive
contributions of the order of several percent from annihilation.
Notice that the annihilation topologies in BBNS lead to
$1/m_b^2$ corrections that are doubly chirally enhanced
(giving rise to the $\rho_A^2$ terms above).

\subsubsection{Scenario 2: Dominance of annihilation topologies}

In another approximation, we assume that 
the non-factorizable part of the $\Delta I = 3/2$ amplitude
is sub-dominant and can be neglected, $r_u(2,3/2)=0$. In addition
we again fix the form factor, $F_0^{B \to \pi} =0.26$. 

From the theoretical point of view, the
situation corresponds to the case where the annihilation graphs
in the BBNS approximation are assumed to be the dominant
source of non-factorizable effect. Alternatively, it can also
be viewed as representing the
case where higher-order contributions from penguin corrections
(``charm'' and ``GIM'' penguins \cite{Ciuchini:1997rj,Buras:1998ra};
for a recent phenomenological fit along these lines,
see \cite{Ciuchini:2004ej}) are the main
source of non-factorizing effects (again, such effects do
not contribute to the amplitudes with $\Delta I=3/2$).

\begin{figure}[tbph]
\small
\begin{tabular}{cc}
\small $r_u(0,1/2)$ & \small $\phi_u(0,1/2)$ \\[-0.3em]
\psfig{file=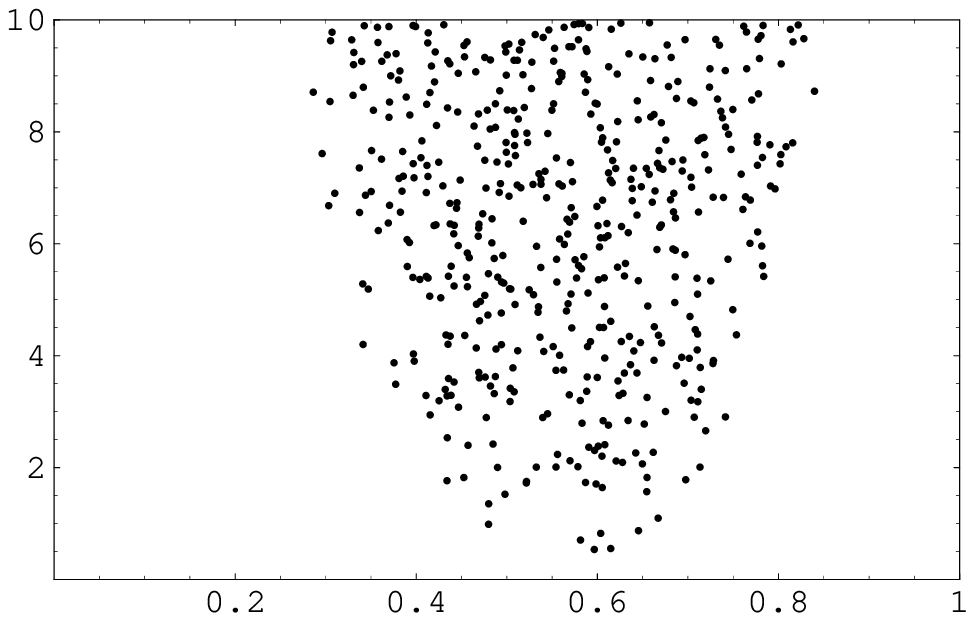, width=0.46\textwidth}
&
\psfig{file=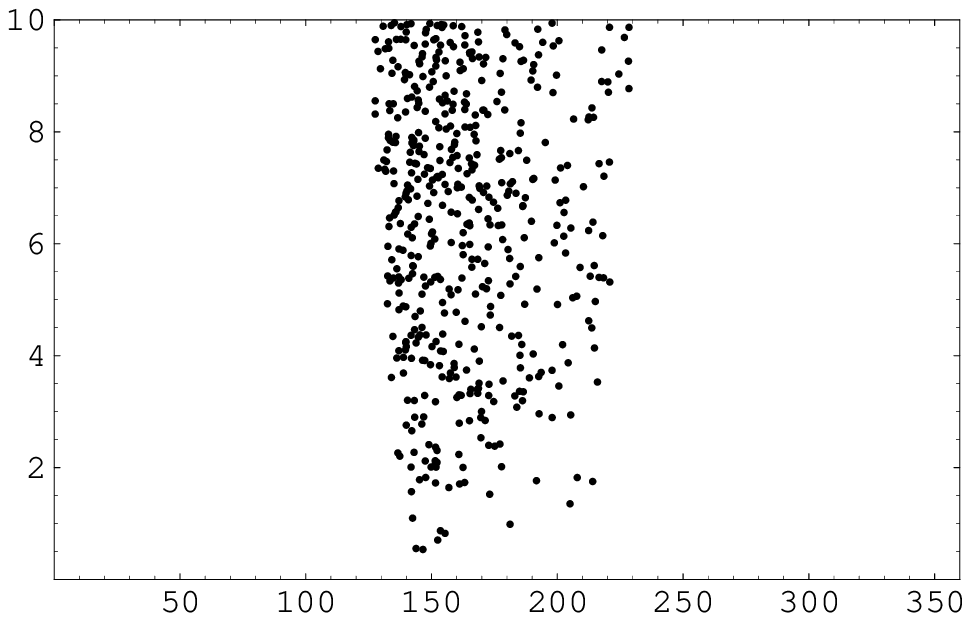, width=0.46\textwidth}
\\[0.1em]
\small $r_c(0,1/2)$ & $\phi_c(0,1/2)$ \\[-0.3em]
\psfig{file=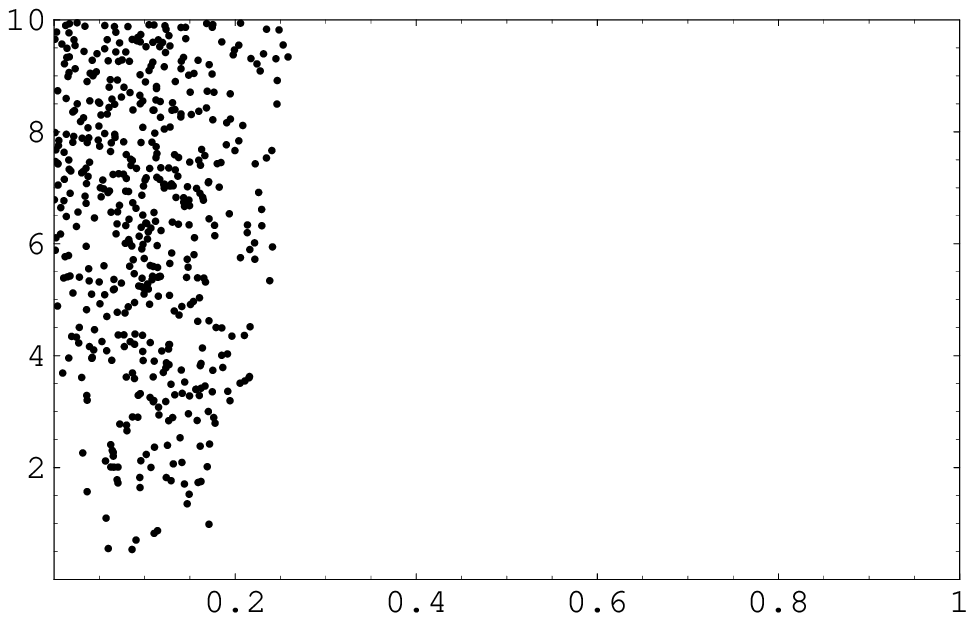, width=0.46\textwidth}
&
\psfig{file=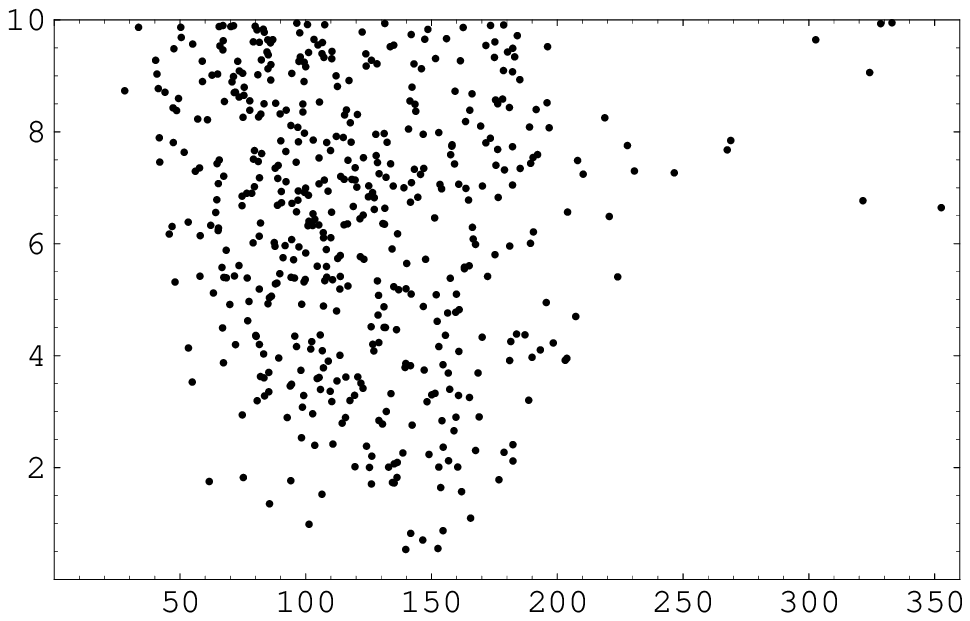, width=0.46\textwidth}
\end{tabular}
\normalsize
\caption{Comparison of $\chi^2$ values for a random
sample of non-factorizable parameter combinations, using
$A_u^{\rm NF}(2,3/2)=0$ and $F_0^{B \to \pi} = 0.26$
(Scenario 2 -- dominance of annihilation/penguins).}
\label{fig:par_scenario2}
\end{figure}

From the plots in Fig.~\ref{fig:par_scenario2} we observe
that a very good description of the data is possible in such
a scenario.
Again, rather large 
non-factorizable effects (i.e.\ 
of the same order as the factorizable ones)
are needed in $A_u(0,1/2)$ and/or  $A_c(0,1/2)$.
The data also show a clear prefererence for 
$A_c^{\rm NF}(0,1/2) < A_u^{\rm NF}(0,1/2)$ in
this scenario.

\subsection{Non-factorizable contributions in SCET}

\label{sec:scet}

Soft-collinear effective theory has been developped as
a systematic tool to study the factorization of different
short- and long-distance modes contributing to inclusive
and exclusive $B$ decays \cite{Bauer:2000yr,Beneke:2002ph,Bauer:2001yt}. In case of
exclusive decays, two kinds of short-distance modes 
are successively integrated out: hard modes (with virtualities
of order $m_b^2$) and hard-collinear modes (with energies of
order $m_b$ and virtualities of order $m_b \Lambda_{\rm QCD}$)
\cite{Hill:2002vw,Bauer:2002aj,Beneke:2003pa}.
The first matching step leads from QCD to the so-called
\scetI. The second matching step leads from
\scetI\ to \scetII. The effective theory
\scetII\ only contains long-distance
modes with virtualities of the order of the QCD scale. In the $B$\/ meson
rest frame these are denoted as ``soft'' (all momentum 
components of order $\Lambda_{\rm QCD}$), and ``collinear''
(one momentum component scales as $m_b$). Fields and
operators in the effective theory have a definite power-counting
in terms of the expansion parameter $\lambda^2=\Lambda_{\rm QCD}/ m_b$.
In this paragraph we will discuss some generic examples for
effective-theory operators that may give rise to 
factorizable and/or non-factorizable effects in 
$B \to \pi\pi$. 

Let us follow the strategy of \cite{Beneke:2003pa} (which has been used
in the context of a factorization proof for the $B \to \pi\ell\nu$
decay) and identify the possible field content of 
effective operators in SCET$_{\rm I}$ that are relevant
for $B \to \pi\pi$. (We will consider light-cone gauge
for the collinear modes, and will drop
soft Wilson lines for simplicity. We also do not explicitely
note Dirac or Lorentz indices.). The two hard-collinear
directions defined by the final-state hadron are denoted
as ``hc'' and ``$\overline{\rm hc}$'', respectively.
It is understood that for all operators that we will list below,
one has a corresponding
term with ${\rm hc} \leftrightarrow \overline{\rm hc}$ interchanged. 
It should also be clear that the different possible
flavour structures of the operators in SCET are obtained
from matching the corresponding operators
in the weak effective hamiltonian by integrating out hard QCD modes
(we will comment on flavor-dependent QED effects below).
 
In the second matching step, one has to generate the
minimal field content 
$$
[\bar q_s \ldots h_v] \, 
[\bar \xi_c \ldots \xi_c] \,
[\bar \xi_{\bar c} \ldots \xi_{\bar c}] 
$$
that is necessary to build up the initial and final state
quantum numbers (the dots stand for additional $q\bar q$ pairs
or gluon fields of the same kind; we do not consider 
decays into flavor-singlet mesons here).
The generation of soft and collinear fields from
hard-collinear ones costs a certain power of the small 
expansion parameter $\lambda$,
which can be read off the corresponding interaction terms in
the SCET$_{\rm I}$ Lagrangian. Examples are \cite{Beneke:2003pa}
\beq
  \xi_{\rm hc} &\stackrel{\lambda}{\rightarrow} & \xi_c \ ,
\cr
  \xi_{\rm hc} &\stackrel{\lambda^2}{\rightarrow} & \xi_c \, A^\perp_{\rm hc}
\ , 
\cr
  A^\perp_{\rm hc} &\stackrel{\lambda}{\rightarrow} & \bar q_s \xi_{\rm hc}
\ , \cr
  A^\perp_{\rm hc} &\stackrel{\lambda^2}{\rightarrow} & \bar q_s \xi_c
\ . \label{chains}
\eeq

\subsubsection{3-body operators}
The minimal possible field content for a SCET$_{\rm I}$
operator, that contributes to $B \to \pi\pi$,  is 
\beq
[\bar \xi_{\rm hc} \, A_{\overline{\rm hc}}^\perp \, h_v] \sim
\lambda^5 \ ,
\label{eq:O8effnaive}
\eeq
where on the right-hand side we indicated the power-counting 
for this operator, following from the SCET$_{\rm I}$ Lagrangian. 
Performing the explicit matching calculation, one finds that
at tree-level the first non-trivial operator is just a copy of
the chromomagnetic term $O_8^g$ in the weak hamiltonian, 
restricted to the particular kinematical situation,\footnote{
The power-counting follows from the leading term
$n_-^\sigma G_{\mu_\perp\sigma} = (n_- \partial) A^\perp_{\rm \overline {hc}}
\sim \lambda$ in the light-cone gauge.}
\beq
&& 
  - C_8^{\rm eff} \, \frac{g_s m_b}{8\pi^2} \,
   \bar \xi_{\rm hc} \, \sigma_{\mu\nu} (1+\gamma_5) \, 
     G^{\mu\nu}_{\rm \overline{hc}} \, h_v
     + ({\rm hc} \to \overline{\rm hc}) 
\ . \label{eq:O8eff}
\eeq
Here $C_8^{\rm eff}$ contains the short-distance contribution
from loop contractions with 4-quark penguin operators.

Starting from the operator in (\ref{eq:O8eff}),
in order to generate the necessary final-state partons,
we need at least two collinear quark fields $\xi_c$ and $\bar \xi_c$
which have to be generated from the hard-collinear field
$\bar \xi_{\rm hc}$, and two collinear quark fields $\xi_{\bar c}$ and
$\bar \xi_{\bar c}$ which should descend from
$A^\perp_{\overline{\rm hc}}$. The first case costs at least a factor
$\lambda^4$, for instance through the chain
\beq
  \bar \xi_{\rm hc} \stackrel{\lambda^2}{\rightarrow}
   \bar \xi_c A_{\rm hc}^\perp
  \stackrel{\lambda^2}{\rightarrow} \bar \xi_c [\bar q_s \xi_c]
 \ . \label{chain:ff}
\eeq
In the second case, we cannot directly produce two collinear
quark fields from the hard-collinear gluon field, because they
would be in a flavor-singlet configuration (the case of
flavor-singlet mesons has been discussed in the context of
QCD factorization in \cite{Beneke:2002jn}). Therefore we need
at least two additional quark fields that do not end up in
the \mbox{corresponding} pion (and therefore have to come from the
initial $B$\/ meson which provides {\it soft}\/ modes). 
A possible branching
\beq
  A_{\rm \bar hc}^\perp \stackrel{\lambda}{\rightarrow}
   \bar q_s \xi_{\overline{\rm hc}}
  \stackrel{\lambda^4}{\rightarrow}
  [\bar q_s \xi_{\bar c}][\bar \xi_{\bar c} q_s]
\label{chain:ns}
\eeq
costs a factor $\lambda^5$ such that the power-counting
for currents in SCET$_{\rm II}$ that descend from 
three-body operators in  SCET$_{\rm I}$ is $\lambda^{14}$.
Together with the power-counting for the hadronic states
the contribution to the $B \to \pi\pi$ amplitude is
$\lambda^7$, which has to be compared with the result
in naive factorization $\sim f_\pi F_0^{B \to \pi} \sim \lambda^5$.
Therefore, contributions from 
three-body operators are $1/m_b$ suppressed
(and therefore do not contribute to the factorization theorem
in the heavy-quark limit).
To obtain a non-vanishing contribution of a 3-body SCET$_{\rm I}$
\mbox{operator} to a $B \to \pi\pi$ matrix element in the diagrammatic
expansion of the BBNS approach, we thus need
at least three gluons, see Fig.~\ref{fig:hardscatt}(b). 
Such diagrams are not considered
in the analysis of power-corrections in \cite{Beneke:1999br},
which has been restricted to one-gluon exchange diagrams.
Up to now, we do not know whether this operator gives factorizable
contributions in the first non-vanishing order in SCET$_{\rm II}$
(i.e.\ $1/m_b$ with respect to the leading contributions from
naive factorization). In any case, from the structure of the
factorization proof for $B \to \pi\ell\nu$ decays in 
\cite{Beneke:2003pa} and similar arguments in \cite{Becher:2003qh},
we expect that, in general, factorization of soft and
collinear modes in SCET$_{\rm II}$ does not hold for power-corrections
obtained in the matching of SCET$_{\rm I}$ to SCET$_{\rm II}$. Therefore, 
at least on the level of $1/m_b^2$ power-corrections,
the operator in (\ref{eq:O8eff}) provides a new source of non-factorizable
corrections. Similarly as for the annihilation diagrams
considered in BBNS, see (\ref{XA}), contributions from
such power-suppressed 8-quark operators in SCET$_{\rm II}$ could be
doubly chirally enhanced. In this case, 
numerically they may be as important as the 
non-factorizable terms included in the BBNS analysis.

\subsubsection{4-body operators}

The leading-power contributions to the function $T_{\rm I}$ 
in (\ref{QCDF}) involve four-quark operators 
in SCET$_{\rm I}$ of the type
\beq
  [\bar \xi_{\rm \overline{\rm hc}} \xi_{\rm hc}][\bar \xi_{\rm hc} h_v]
 \sim \lambda^6 \ .
\eeq
The conversion of the hard-collinear quark pair to a collinear one
costs twice a power of $\lambda$, and the conversion of 
$\bar \xi_{\rm \overline{\rm hc}}$ via (\ref{chain:ff})
costs a factor of $\lambda^4$. Therefore,
these four-quark operators match onto 6-quark operators in
\scetII\ which scale as $\lambda^{12}$ and, according to
the discussion in the previous paragraph, are leading power.

Another 4-quark operator that is allowed in \scetI \,by  momentum
conservation is of the form
\beq
  [\bar q_s h_v][\bar \xi_{\rm \overline{\rm hc}} \xi_{\rm hc}] \sim \lambda^8
\ .
\label{eq:anni}
\eeq
The conversion of hard-collinear quark fields into at least
two collinear fields via (\ref{chain:ff})
costs each a factor of $\lambda^4$. Therefore,
these operators match onto 8-quark operators in \scetII\ which
scale at least as $\lambda^{16}$ and are thus $1/m_b^2$ suppressed. 
This corresponds to an annihilation topology which involves a
higher Fock state $(b\bar q)(q'\bar q')$ in the $B$ meson.
Notice that, again, the suppression by $1/m_b^2$ can be numerically
compensated by {\it two}\/ chiral enhancement factors coming
from the wave functions of the two final-state mesons.

We may also consider 4-body operators that involve
additional gluons, like
\beq
  [A^\perp_{\rm \overline{\rm hc}}][\bar \xi_{\rm hc} A^\perp_{\rm hc} h_v]
\sim \lambda^6 \quad
\stackrel{{\rm SCET}_{\rm II}}{\rightarrow}
\quad
 \int \ldots [\bar q_s \xi_{\bar c}][\bar \xi_{\bar c} q_s]
  [\bar q_s \xi_c][\bar \xi_c h_v] 
  \sim \lambda^{14} \ ,
\eeq
or
\beq
  [A^\perp_{s} A^\perp_{\rm hc}][\bar \xi_{\overline{\rm hc}} h_v]
\sim \lambda^7\quad
\stackrel{{\rm SCET}_{\rm II}}{\rightarrow}
\quad
 \int \ldots A^\perp_s [\bar q_s \xi_{\bar c}][\bar \xi_{\bar c} q_s]
  [\bar q_s \xi_c][\bar \xi_c h_v] 
  \sim \lambda^{16} \ ,
\label{4bodypeng}
\eeq
where again we have to use (\ref{chain:ns}) to obtain 
flavour non-singlet collinear quark configurations from
a single hard-collinear gluon field. 

\subsubsection{Remark on the treatment of charm quarks}

\label{sec:charmq}

One may also worry about 4-quark operators involving
charm quarks. In the BBNS approach, the charm quarks are treated
as {\it hard}\/ modes (i.e.\ $m_c = {\cal O}(m_b)$), 
and therefore they are integrated out in the first matching
step and do not appear as degrees of freedom in \scetI.
Alternatively, one may take the point of view that 
$m_c^2 \sim \Lambda m_b$, i.e.\ $m_c \ll m_b$ in the heavy
quark limit. Still, since charm quarks do not appear as
external partonic degrees of freedom in charmless non-leptonic
$B$\/ decays, they cannot induce endpoint singularities,
and can, in any case, be treated perturbatively. 
The effect of the alternative power-counting scheme
merely amounts to expanding the
hard coefficient functions in terms of $m_c/m_b$, which 
corresponds to integrating out the charm quarks in the
second matching step \scetI~$\to$~\scetII. Via the 
renormalization group running within \scetI\ one would
also resum logarithms $\ln m_c/m_b$. Numerically, the
effect onto factorizable amplitudes should be marginal.
In the non-factorizable contributions, on the other hand,
the crucial effects come from the endpoint divergences
and chiral enhancement related to {\it light quarks}\/,
and not from charm quarks, if $m_c \gg \Lambda_{\rm QCD}$.

In \cite{Bauer:2004tj} it has been argued 
that one should also
consider the possibility of non-relativistic charm and gluon modes
in SCET$_{\rm I}$.
This would correspond to operators of the type
$$
  [\bar \xi_{\overline{\rm hc}} h_v ][\bar c c]_{\rm NR} \ ,
$$
where the invariant mass of the $c\bar c$ pair happens
to be close to $4 m_c^2$ which corresponds to a light-cone
momentum fraction for the quark field
$\bar \xi_{\overline{\rm hc}}$ 
of the order $u=1-4m_c^2/m_b^2 \simeq 0.6$.
However, since charm modes do not appear as external degrees
of freedom, hadronic matrix elements of the above operator 
in SCET$_{\rm I} \times $~NRQCD would vanish.
One may ask the question, where the effect of 
charm resonances (i.e.\ non-relativistic $c\bar c$ bound states,
$J/\psi, \psi', \ldots$) would show up in the effective theory
framework. As explained above, as charm quarks do not appear as
external states, they can be formally integrated out, resulting in
a quark determinant with charm quarks in the background of soft
and collinear gluon fields. The treatment of charm quarks is
thus fully inclusive, and the appearance of charm resonances
resembles the well-known cases of $e^+e^- \to \mbox{hadrons}$ or
$B \to X_s \ell^+\ell^-$: 
When integrating the invariant-mass
spectrum over a large enough region, the
effect of charm resonances provides power-corrections to
the inclusive rate. 

In our case, the pion distribution amplitude serves as the ``detector''
with a \mbox{``sensitivity''} $\phi_\pi(u = 1-q^2/m_b^2)$.
Integrating over all momentum fractions $u$, the effect of
charm resonances translates into power-corrections, which should
be attributed to the matching coefficients of
sub-leading operators in SCET$_{\rm I}$. 
One can even perform a numerical estimate
of such charm-resonance effects by
using the same phenomenological treatment
as for 
$b \to s \ell^+ \ell^-$ 
\cite{Ali:1991is,Kruger:1996cv}.
We find
that the standard pion distribution amplitude is sufficiently
broad to wash out the effect of exclusive charm resonances.
Therefore it is not clear to us  
in what sense the inclusion of non-relativistic modes
in SCET$_{\rm I}$ should lead to an enhancement of charm
penguin contributions in non-leptonic $B$ decays, as has
been argued in \cite{Bauer:2004tj}. Also in a recent
analysis within QCD sum rules \cite{Khodjamirian:2003eq}, 
an unnatural enhancement
of charm-penguin contributions to non-leptonic $B$ decays
is not observed, and the perturbative treatment of charm
quarks seems to be justified.

\subsubsection{5-body operators}

Two examples of
4-quark operators with an additional
hard-collinear gluon are
\beq
  [\bar \xi_{\rm hc} \xi_{\overline{\rm hc}}]
  [\bar \xi_{\overline{\rm hc}} A_{\rm hc}^\perp h_v] &\sim&  \lambda^7 \ ,
\\[0.2em]
  [\bar \xi_{\rm hc} A_{\rm hc}^\perp  \xi_{\rm hc}]
  [\bar \xi_{\overline{\rm hc}} h_v] &\sim&  \lambda^7 \ .
\eeq
With similar arguments as above,
the first term matches onto a leading-power 
6-quark operator in \scetII, which corresponds to the
factorizable hard-spectator diagrams in the
QCD factorization approach. The second term gives
a $1/m_b$ power-suppressed contribution, which serves
as one of the candidates to compensate the endpoint divergences
parametrized by $X_H$. 
There are a number of further 5-body structures, which
we are not going to discuss in detail here.

\subsubsection{6-body operators}

The simplest 6-quark operator in \scetI\ has the form
\beq
  [\bar \xi_{\rm hc} \xi_{\overline{\rm hc}}]
  [\bar \xi_{\overline{\rm hc}} \xi_{\rm hc}]
  [\bar q_s h_v] \sim \lambda^{10}
\eeq
The leading-power matching onto \scetII\ gives a 6-quark operator
of the order $\lambda^{14}$  which corresponds
to the $1/m_b$\/-suppressed annihilation graphs in
the BBNS approach.

Another possibility is a 4-quark operator with two
additional transverse gluons, e.g.\
\beq
  [\bar \xi_{\rm hc} \xi_{\overline{\rm hc}}][\bar q_s h_v]
  A_{\rm hc}^\perp A_{\overline{\rm hc}}^\perp 
  \sim \lambda^{10}
\eeq
which amounts to annihilation with additional
radiation of hard-collinear gluons. After matching
onto \scetII, the minimal quark content refers to
an 8-quark operator which scales as $\lambda^{16}$,
and is therefore $1/m_b^2$ suppressed with respect
to the leading-power contributions, and should be
related to the terms of order $X_A^2$ in (\ref{XA}).

\subsubsection{Summary: 
Factorizable and non-factorizable SCET operators}

We summarize the examples for \scetI\ and \scetII\ operators
in Table~\ref{tab:scet}. The identification of the
power-corrections as non-factorizable is tentative 
(without a more detailed analysis, which is beyond the scope of 
this work, we cannot exclude that some of the operators
identified as non-factorizable at a certain sub-leading power
in $1/m_b$ are actually free of endpoint divergences in \scetII).
The factorization of the two structures contributing at
leading power (as indicated by $\surd$), can be understood
by applying the same rules as in the $B \to \pi\ell\nu$ case
\cite{Beneke:2003pa} (this point has also been realized in 
\cite{Bauer:2004tj}).

Notice that another set of possible operators is obtained
by changing one gluon field into a photon field. The 
contributions of such operators are suppressed by the
ratio of electromagnetic and strong coupling constants,
$\alpha_{\rm em}/\alpha_s$. They may be
important in isospin-breaking observables for $B \to \pi K$
decays, if they enter with large Wilson coefficients and
large CKM elements, and if the $1/m_b$ suppression is compensated
by large numerical factors.

\begin{table}[hbt]
\begin{center}
\begin{tabular}{|c | c ||c | c||c || c |}
  \hline
&&&&&
\\[-0.9em]
\scetI & power & $A^{\rm F}$ & $A^{\rm NF}$ & \scetII  & BBNS
\\[0.2em]
\hline\hline
&&&&&
\\[-0.9em]
$[\bar \xi_{\rm hc} \, A^\perp_{\overline{\rm hc}} \, h_v]$ 
& $   \lambda^5$
& -- & $1/m_b$ & 
$[\bar q_s \bar q_s .. q_s h_v]
 [\bar \xi_{\bar c} .. \xi_{\bar c}][\bar \xi_c .. \xi_c]$
&
\\[0.2em]
\hline
&&&&&
\\[-0.9em]
$[\bar \xi_{\rm \overline{\rm hc}} \xi_{\rm hc}][\bar \xi_{\rm hc} h_v]$
& $   \lambda^6$
& $\surd$ & $1/m_b$ &
$[\bar q_s .. h_v]
 [\bar \xi_{\bar c} .. \xi_{\bar c}][\bar \xi_c .. \xi_c]$
& in $T_{\rm I}$
\\[0.2em]
\hline
&&&&&
\\[-0.9em]
$[A^\perp_{\rm \overline{\rm hc}}][\bar \xi_{\rm hc} A^\perp_{\rm hc} h_v]$
& $  \lambda^6$
& -- & $1/m_b$ &
$[\bar q_s \bar q_s .. q_s h_v]
 [\bar \xi_{\bar c} .. \xi_{\bar c}][\bar \xi_c .. \xi_c]$
&
\\[0.2em]
\hline
&&&&&
\\[-0.9em]
$ [\bar \xi_{\rm hc} \xi_{\overline{\rm hc}}]
  [\bar \xi_{\overline{\rm hc}} A_{\rm hc}^\perp h_v]$
& $   \lambda^7$
& $\surd $ & $1/m_b$
& $[\bar q_s .. h_v]
 [\bar \xi_{\bar c} .. \xi_{\bar c}][\bar \xi_c .. \xi_c]$
& in $T_{\rm II}$, $X_H$
\\[0.2em]
\hline
&&&&&
\\[-0.9em]
$ [\bar \xi_{\rm hc} A_{\rm hc}^\perp  \xi_{\rm hc}]
  [\bar \xi_{\overline{\rm hc}} h_v]$
& $   \lambda^7$ 
& -- & $1/m_b$ &
 $[\bar q_s .. h_v]
 [\bar \xi_{\bar c} .. \xi_{\bar c}][\bar \xi_c .. A_c^\perp \xi_c]$
&
\\[0.2em]
\hline
&&&&&
\\[-0.9em]
$[A^\perp_{s} A^\perp_{\rm hc}][\bar \xi_{\overline{\rm hc}} h_v]$
& $  \lambda^7$
& -- & $1/m_b^2$ &
$[\bar q_s \bar q_s .. A_s^\perp q_s h_v]
 [\bar \xi_{\bar c} .. \xi_{\bar c}][\bar \xi_c .. \xi_c]$
&
\\[0.2em]
\hline
&&&&&
\\[-0.9em] 
$[\bar q_s h_v][\bar \xi_{\rm \overline{\rm hc}} \xi_{\rm hc}]$
& $  \lambda^8$
& -- & $1/m_b^2$ &
$[\bar q_s \bar q_s .. q_s h_v]
 [\bar \xi_{\bar c} .. \xi_{\bar c}][\bar \xi_c .. \xi_c]$
&
\\[0.2em]
\hline
&&&&&
\\[-0.9em] 
$ [\bar \xi_{\rm hc} \xi_{\overline{\rm hc}}]
  [\bar \xi_{\overline{\rm hc}} \xi_{\rm hc}]
  [\bar q_s h_v]$
& $   \lambda^{10}$
& -- & $1/m_b$ 
& $[\bar q_s .. h_v]
 [\bar \xi_{\bar c} .. \xi_{\bar c}][\bar \xi_c .. \xi_c]$
& in $X_A$
\\[0.2em]
\hline
&&&&&
\\[-0.9em] 
$ [\bar \xi_{\rm hc} \xi_{\overline{\rm hc}}][\bar q_s h_v]
  A_{\rm hc}^\perp A_{\overline{\rm hc}}^\perp $
& $ \lambda^{10}$
& -- & $1/m_b^2$ 
& $[\bar q_s \bar q_s .. q_s h_v]
 [\bar \xi_{\bar c} .. \xi_{\bar c}][\bar \xi_c .. \xi_c]$
& 
\\[0.2em]
\hline
&&&&&
\\[-0.9em] 
\ldots & \ldots & \ldots & \ldots & \ldots & \ldots
\\[0.2em]
\hline
\end{tabular}
\end{center}
\caption{\label{tab:scet} An incomplete list of the field content of
SCET$_{\rm I}$ operators, contributing to non-leptonic $B$\/ decays
into light hadrons. See text for details.
}
\end{table}

\subsection{Non-factorizable effects from 
long-distance penguins}

\begin{figure}[tbph]
\begin{center}
(a)
\psfig{file=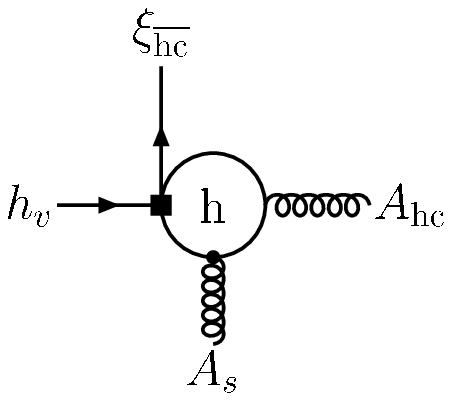, height=0.2\textwidth}
\hspace{3em}
(b)
\psfig{file=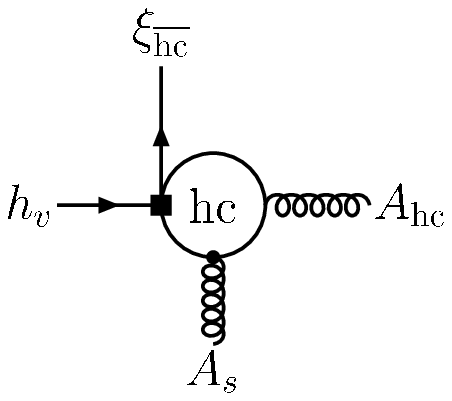, height=0.2\textwidth}
\hspace{2.5em}
\psfig{file=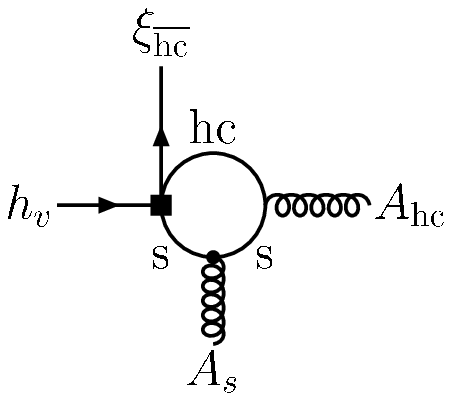, height=0.2\textwidth}
\end{center}
\caption{Examples for (a) short- and
(b) long-distance penguin contributions to 
a 4-body Greens function in SCET$_{\rm I}$ via hard, 
hard-collinear or soft quark loop.}
\label{fig:penghc}
\end{figure}

It is often argued 
that a main source of non-factorizable
effects in $B \to PP$ decays should be attributed to
long-distance penguin topologies \cite{Ciuchini:1997rj,Buras:1998ra},
where two quarks of one of
the four-quark operators ${\cal O}_{1-6}$ in the electroweak hamiltonian
are contracted to a loop which radiates one or more gluons.
As explained above, in the effective theory approach, long-distance
modes in SCET$_{\rm I}$ are hard-collinear and soft quark and gluon
fields, i.e.\ the effect of $b$ (and usually also $c$) quarks is
already absorbed into matching coefficients,
see Fig.~\ref{fig:penghc}(a). Examples for
long-distance penguin diagrams in 
SCET$_{\rm I}$ are  shown in Fig.~\ref{fig:penghc}(b).
They reflect contributions to one of the 4-body operators 
(\ref{eq:anni}) discussed above, 
which gives power-suppressed contributions to the 
decay amplitudes. 
Remember  that above, we identified this type of operator as the source for
annihilation contributions.
Actually, in the non-perturbative region the diagrammatic language in
terms of soft quark and gluon propagators is not appropriate anymore,
and the distinction between annihilation and penguin contractions
is not clear cut. On the other hand, in SCET one has an additional
correlation between flavour quantum numbers and the type of fields
(hard-collinear or soft). For the above example, the operators
\beq
  [\bar q_s h_v] [\bar \xi_{\overline{\rm hc}} \xi_{\rm hc}]_{I=0}
  \qquad \mbox{and} \qquad
  [\bar q_s h_v] [\bar \xi_{\overline{\rm hc}} \xi_{\rm hc}]_{I=1} \ ,
\eeq
where the total isospin of the two hard-collinear fields is
different, can be distinguished. Notice however that, in general,
interactions between
soft and hard-collinear quarks can lead to
mixing of the two operators in SCET$_{\rm I}$.
Notice that by momentum conservation
both hard-collinear fields are in an endpoint configuration
\beq
  p_{\rm hc}^\mu &=& E n_-^\mu + \Delta p^\mu, 
  \qquad |\Delta p^\mu| \sim \Lambda\cr
  p_{\overline{\rm hc}}^\mu &=& E n_+^\mu + \Delta \bar p^\mu, \qquad |\Delta \bar p^\mu| 
   \sim \Lambda \ ,
\eeq
where $E\simeq m_b/2$ denotes the pion energy in the $B$\/ meson rest frame,
and $n_-$ and $n_+$ are light-cone vectors which satisfy
$n_-^2=n_+^2=0$, $n_- n_+=2$.
The invariant mass of the gluon pair is of order $\sqrt{\Lambda m_b}$,
and therefore the light-quark thresholds in the
loop give rise to an imaginary part in the soft loop. 
We emphasize that  for this configuration the factors of $\alpha_s$
belonging to the soft gluon and to the endpoint-gluon do not
count as perturbative. Therefore the latter  reflects a mechanism to
generate non-perturbative strong phases. On the other hand,
as already mentioned, this contribution is power-suppressed by
at least $1/m_b^2$, in accordance with the general arguments
in \cite{Donoghue:1996hz} and \cite{Beneke:1999br}, and, in principle,
the endpoint-configuration should be suppressed by Sudakov
form factors. On the other hand, we could have a sizeable numerical
enhancement from chiral factors, but this cannot be quantified 
in a satisfactory  way at the moment.

\subsubsection{Scenario 3: Dominance of strong phases 
in $A_c(0,1/2)$}

Sometimes the main source of non-factorizable
effects in $B \to \pi\pi$ decays is attributed to
matrix elements of the operators 
${\cal O}_{1,2}^c \sim [\bar d b][\bar c c]$ in the electroweak
hamiltonian (see e.g.\ \cite{Ciuchini:1997rj,Buras:1998ra}
for the general argument, and \cite{Barshay:2004ra} for a recent
hadronic model). This would correspond to 
long-distance penguin topologies with charm quarks in the loop.
In the effective theory approach this would require to treat 
$4m_c^2 \leq \Lambda m_b$.
Also, the line-of-reasoning in \cite{Bauer:2004tj} (see the discussion
about non-relativistic charm modes, above)  has lead the
authors to conclude that charm penguins are the primary source of
strong rescattering phases. 

\begin{figure}[tbph]
\small
\begin{tabular}{cc}
\small $r_u(0,1/2)$ & \small $r_u(2,3/2)$ \\[-0.3em]
\psfig{file=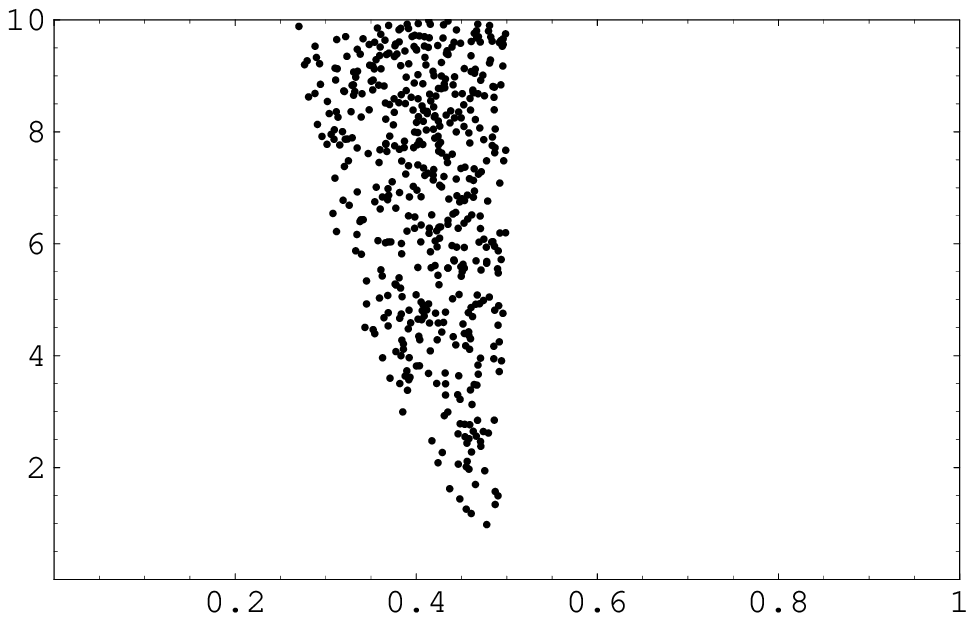, width=0.46\textwidth}
&
\psfig{file=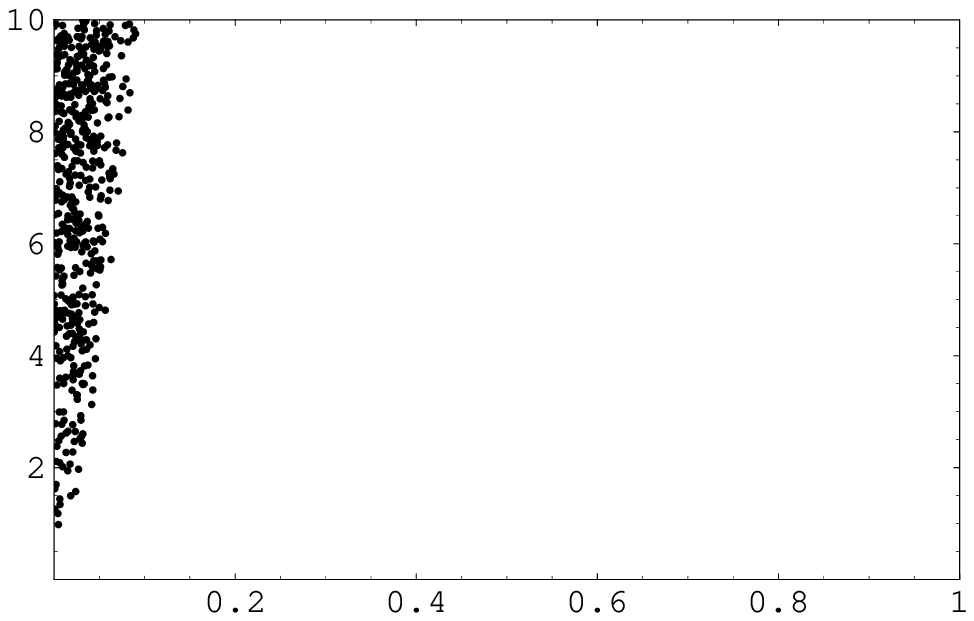, width=0.46\textwidth}
\\[0.1em]
\small $r_c(0,1/2)$ & $\phi_c(0,1/2)$ \\[-0.3em]
\psfig{file=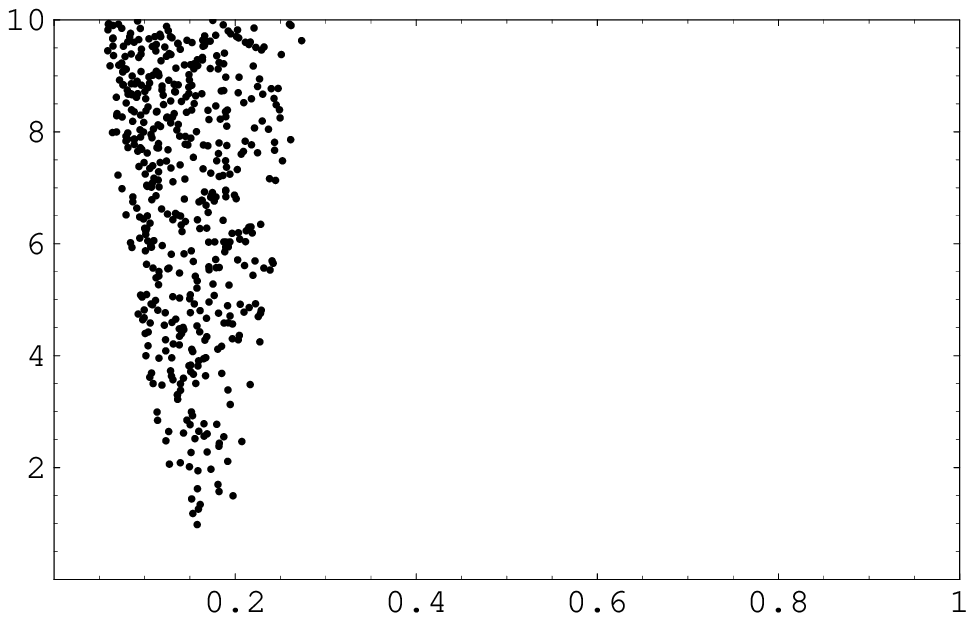, width=0.46\textwidth}
&
\psfig{file=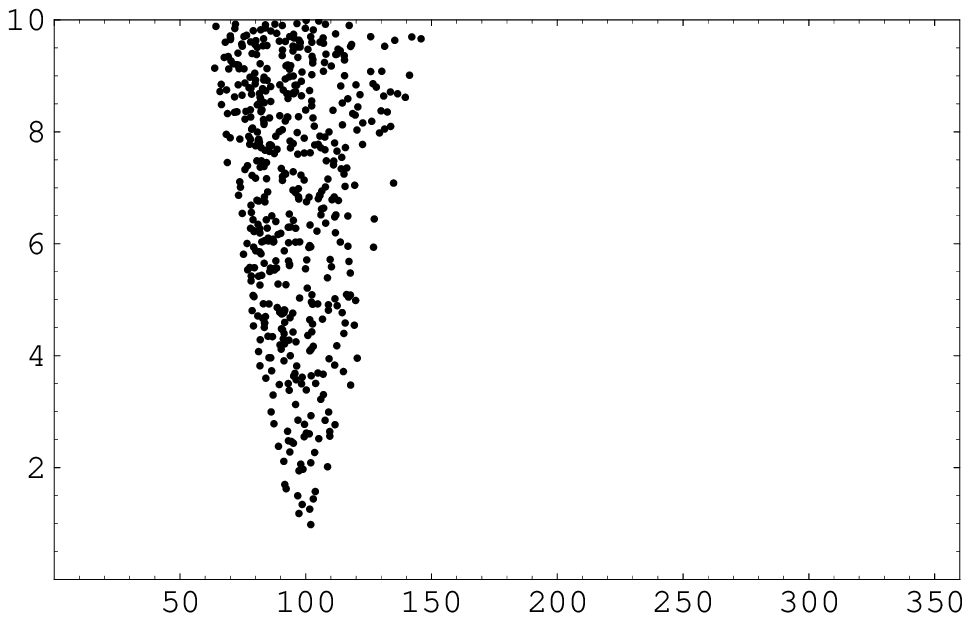, width=0.46\textwidth}
\end{tabular}
\normalsize
\caption{Comparison of $\chi^2$ values for a random
sample of non-factorizable parameter combinations, using
$\phi_u(0,1/2)=\phi_u(2,3/2)=180^\circ$,
$F_0^{B \to \pi}=0.26$, $r_u(0,1/2)<0.5$, $r_u(2,3/2)<0.2$ 
(Scenario 3 -- dominance of strong phases from $A_c(0,1/2)$). 
}
\label{fig:par_scenario3}
\end{figure}

In a more general setup, such a situation 
can be simulated by setting the phases $\phi_u(0,1/2),
\phi_u(2,3/2)= 180^\circ$. Furthermore, we require
moderate values for the non-factorizable parameters,
$r_u(0,1/2) < 0.5$ and $r_u(2,3/2)<0.2$. Notice that, in general,
one cannot decide whether the remaining contributions
to $A_c^{\rm NF}(0,1/2)$ are to be identified as long-distance
penguin or annihilation topologies (and therefore also
the phenomenological discussion in \cite{Pham:2004xw} is covered).
The result of this scenario (where the value of the form factor
is again fixed as $0.26$) is shown in Fig.~\ref{fig:par_scenario3}.
We observe that the scenario is very restrictive, 
leading to $r_c(0,1/2)\simeq 0.15$, and
$\phi_c(0,1/2) \simeq 100^\circ$.
This implies that, based on phenomenological assumptions,
the theoretical predictivity compared to the more general
case in scenario~2 is improved. However, 
these constraints, used in a CKM analysis, probably represent
a large theoretical bias.

We should mention  that the phenomenological
scenario constructed in \cite{Bauer:2004tj}
differs from the one discussed here in two points: 
(i) In \cite{Bauer:2004tj} the remaining contributions to $r_u(I,\Delta I)$ 
are attributed to {\it factorizable}\/
hard-scattering contributions, where the relative weight for
different isospin amplitudes is fixed but the absolute
normalization is kept arbitrary. (ii) The form factor value is left
as a free parameter. In this case the authors found a very small
value $F_0^{B \to \pi}\simeq 0.17$ which seems to contradict 
the findings from
QCD sum rules and would point to large hard-scattering corrections
relative to the naively factorizing terms.
We understand this as an indication that the 
neglect of other non-factorizable corrections in \cite{Bauer:2004tj} 
is questionable. Since the transition form factor itself and the
factorizable corrections actually represent one part of the calculation
where we have some theoretical control on the parametric errors,
we believe that one should stick to the present theoretical estimates,
unless a more direct experimental determination of these terms
(from $B \to \pi \ell\nu$ and $B \to \gamma\ell\nu$) gives more
reliable numbers.

What remains true is  that
the assumption about the dominance of
strong rescattering phases in the amplitude $A_c (0,1/2)$
cannot be excluded with present data. In view of the
already mentioned QCD sum rules result \cite{Khodjamirian:2003eq}
and the general arguments in Section~\ref{sec:charmq}, it does
not seem very plausible that these effects are due to charm
penguins alone.

\subsubsection{Scenario 4: Equal strong phases in
$A_u(0,1/2)$ and $A_c(0,1/2)$}

As explained above, in the standard framework,
long-distance penguins would only involve light quarks.
Imaginary parts are related to the soft momentum regions
in these light-quark loops. If this was the leading
mechanism to generate strong phases, one may indeed expect
that the corresponding contributions to $A_u(0,1/2)$ and
$A_c(0,1/2)$ differ in moduli, due to different 
Wilson coefficients, and short-distance coefficient functions
in SCET, but yield the same strong phases
$\phi_u(0,1/2)=\phi_c(0,1/2)$.
(More precisely, this situation is always realized
if the non-factorizable contributions are dominated by
only one operator in the $\Delta I = 1/2$ 
SCET$_{\rm I}$ hamiltonian, e.g.\ the operator 
in (\ref{eq:anni}). Strong phases can be induced 
either by soft quark loops in penguin diagrams, 
or by soft gluon rescattering in annihilation diagrams. 
As already mentioned, to disentangle the two possibilities
does not make sense for non-factorizable operators.)
The corresponding 4-quark operators with the largest
Wilson coefficients are ${\cal O}_{1,2}^u$, and therefore
it is also conceivable that  $r_u(0,1/2)>r_c(0,1/2)$.
As one can see from the general set-up in
scenario~2 (see Fig.~\ref{fig:par_scenario2}), both
these assumptions seem to be in line with experiment.

\begin{figure}[tbph]
\small
\begin{center}
\begin{tabular}{c}
 $r_u(0,1/2)$ \\[-0.3em]
\psfig{file=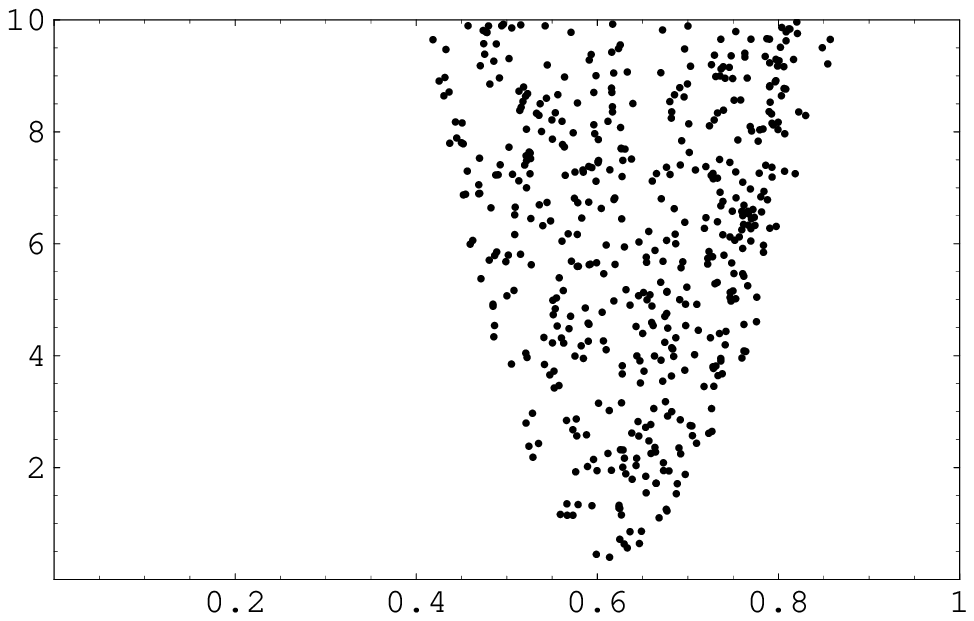, width=0.46\textwidth}
\\[0.1em]
$r_c(0,1/2)$  \\[-0.3em]
\psfig{file=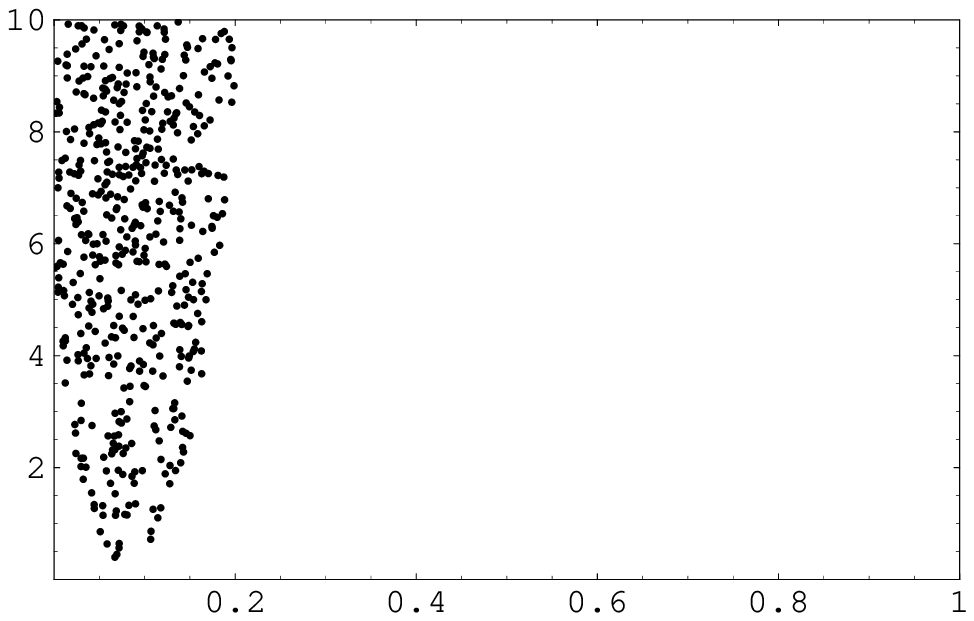, width=0.46\textwidth}
\\[0.1em]
$\phi_u(0,1/2)$ \\[-0.3em]
\psfig{file=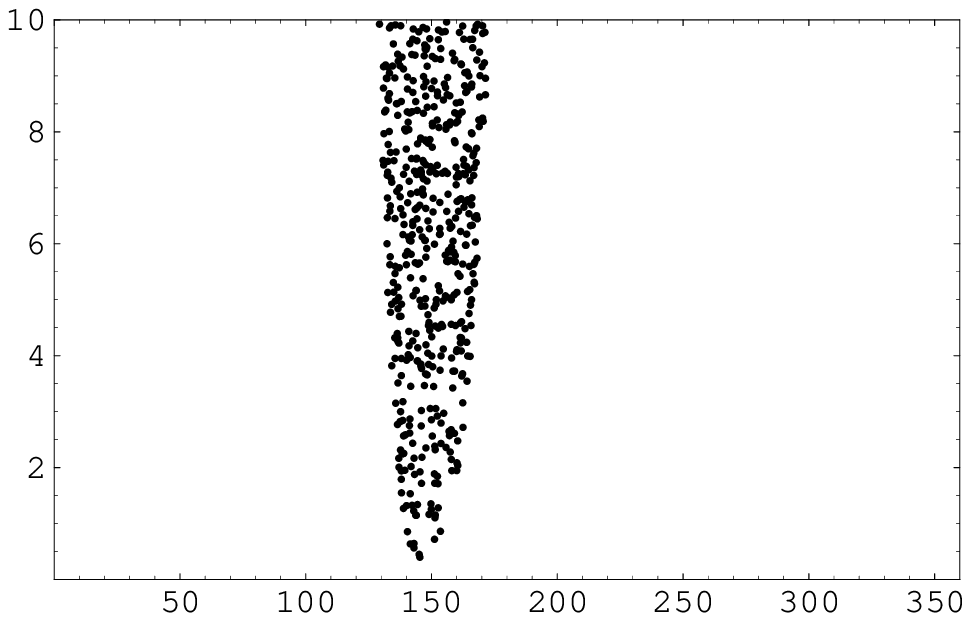, width=0.46\textwidth}
\end{tabular}
\end{center}
\normalsize
\caption{Comparison of $\chi^2$ values for a random
sample of non-factorizable parameter combinations, using
$F_0^{B \to \pi}=0.26$, 
$\phi_u(0,1/2)=\phi_c(0,1/2)$ and
$r_u(2,3/2)=0$ (Scenario 4 -- equal phases for $A_u^{\rm NF}(0,1/2)$,
and $A_c^{\rm NF}(0,1/2)$).}
\label{fig:par_scenario4}
\end{figure}

We therefore define another scenario~4 with four constraints, 
and thus only three free parameters,
$A_u^{\rm NF}(2,3/2)=0$, 
$\phi_c(0,1/2)=\phi_u(0,1/2)$,
and $F_0^{B \to \pi}=0.26$.
The comparison with the experimental data
is shown in Fig.~\ref{fig:par_scenario4}.
Not surprisingly, the largest restriction concerns the value 
of $\phi_u(0,1/2)=\phi_c(0,1/2)$ which is tuned
to values around  $150^\circ$.
But the values of $r_u(0,1/2)$ and $r_c(0,1/2)$
which  lead to a good description of the data are still rather
generic, $r_u(0,1/2) \simeq 0.4-0.8$ and
$r_c(0,1/2) < 0.2$.

\subsection{Lessons from $B \to \pi\pi$}

We conclude: 
 \begin{itemize}
    \item The present data on $B \to \pi\pi$ decays require
      non-vanishing non-factorizable corrections which should
      be related to $1/m_b$ corrections in the factorization
      formula~(\ref{QCDF}). (Here we assume that the
      CKM elements are given by their SM fit values.)

    \item The dynamical origin of these corrections remains
      a theoretical challenge, and different phenomenological
      assumptions can accomodate the data. This includes
      scenarios in the spirit of BBNS,
      where non-factorizable corrections can still
      be moderate for certain hadronic input parameters. 
      On the other hand, the central values
      of experimental data seem to point to rather large
      non-factorizable contributions, in particular for
      the isospin amplitude $A_u(0,1/2)$, which can only
      be reached by pushing the hadronic parameters in 
      the BBNS approach to the limits.

    \item Different assumptions about the dominance of certain
    decay topologies are consistent with the data. 
    However, the additional assumptions and constraints may
    lead to a strong theoretical bias when used in CKM fits.
    Whereas it seems safe to neglect the non-factorizable
    contributions to the isospin amplitude $A_u(2,3/2)$, in
    general, both large contributions to $A_u(0,1/2)$ and
    $A_c(0,1/2)$ should be taken into consideration. 
    They may be related to either long-distance 
    penguin or annihilation topologies which correspond to
    matrix elements of power-suppressed operators in 
    soft-collinear effective theory. 

    \item The picture of large non-factorizable penguin 
     contributions still appears to be attractive 
     from the phenomenological point of view. Although
     they formally appear on the level of power-corrections,
     their numerical impact remains a matter of debate. We have
     given some arguments  that in the effective theory
     framework, one would need
     to consider more complicated diagrams than those
     taken into account in \cite{Beneke:1999br} 
     or \cite{Bauer:2004tj}, in order to
     cover such effects. We also saw that the distinction
     between annihilation and penguin topologies is not
     clear cut for non-factorizable contributions.
     We also find it unlikely  that large
     non-factorizable effects to
     $A_c(0,1/2)$ are coming from charm-quark loops. 

    \item In the extreme case, we may 
     assume that the non-factorizable effects required
     by experimental data are dominated by one 
     non-factorizable operator in the $\Delta I=1/2$ 
     SCET$_{\rm I}$ Lagrangian, leading to equal strong phases
     for $A_u^{\rm NF}(0,1/2)$ and $A_c^{\rm NF}(0,1/2)$.
     Comparison with experimental data shows that 
     these long-distance contributions to the decay amplitudes,
     are about an order of magnitude larger than the factorizable penguin
     contributions contained in (\ref{QCDF}). We note
     that in the BBNS approach this
     scenario can be simulated by allowing for significantly 
     larger non-factorizable annihilation contributions $X_A$.

\end{itemize}

\section{Some Remarks on  $B \rightarrow K \pi$ Decays}

\label{sec:piK}

In the future, we expect new and/or more accurate 
data within the whole class of $B \to PP $  decays 
from the $B$ factories and hadron machines. 
This is crucial for a better understanding of the strong dynamics in
charmless non-leptonic $B$ decays.
In this work, we limit our analysis to some general remarks. 
We plan to give a comprehensive analysis of 
$B \rightarrow K \pi$ decays in a forthcoming paper.

\subsection{Isospin decomposition for $B \to \pi K$}

The isospin decomposition of $B \to \pi K$ amplitudes reads
\beq
 \langle \pi^- \bar K^0 
    | H_{\rm eff} | B^- \rangle & \simeq & 
     \lambda_u^{(s)} \left[ A_u(1/2,0) +A_u(1/2,1)+ A_u(3/2,1) \right]
+ \cr &&\lambda_c^{(s)} \left[ A_c(1/2,0) +A_c(1/2,1)+ A_c(3/2,1) \right]
\ , \\[0.3em]
 \sqrt2 \,
 \langle \pi^0 K^- 
    | H_{\rm eff} | B^- \rangle & \simeq & 
     \lambda_u^{(s)}  \left[ -A_u(1/2,0) - A_u(1/2,1)+ 2 A_u(3/2,1) \right]
    +\cr &&
     \lambda_c^{(s)}  \left[ -A_c(1/2,0) - A_c(1/2,1)+ 2 A_c(3/2,1) \right]
\ , \\[0.3em]
\langle \pi^+ K^- 
    | H_{\rm eff} | \bar B^0 \rangle & \simeq & 
     \lambda_u^{(s)}  \left[ -A_u(1/2,0) + A_u(1/2,1)+  A_u(3/2,1) \right]
    +\cr &&
     \lambda_c^{(s)}  \left[ -A_c(1/2,0) + A_c(1/2,1)+  A_c(3/2,1) \right]
\ , \\[0.3em]
\sqrt2 \,
 \langle \pi^0 \bar K^0 
    | H_{\rm eff} | \bar B^0 \rangle & \simeq & 
     \lambda_u^{(s)}  \left[ A_u(1/2,0) - A_u(1/2,1)+  2 A_u(3/2,1) \right]
    +\cr &&
     \lambda_c^{(s)}  \left[ A_c(1/2,0) - A_c(1/2,1)+  2 A_c(3/2,1) \right]
\ , 
\eeq
reflecting the isospin relation
\beq
 \langle \pi^- \bar K^0 | H_{\rm eff} | B^- \rangle
+ \sqrt2 \, \langle \pi^0 K^- | H_{\rm eff} |  B^- \rangle
- \langle \pi^+ K^-  | H_{\rm eff} | \bar B^0 \rangle 
- \sqrt2 \,\langle \pi^0 \bar K^0 | H_{\rm eff} | \bar B^0 \rangle 
&=& 0 \ .
\cr &&
\eeq
For comparison, we also quote the connection with the
parametrization used in \cite{Beneke:1999br}
\beq
  P &=& \lambda_c^{(s)} 
  \left( A_c(1/2,0) + A_c(1/2,1) + A_c(3/2,1) \right)
\ ,\\[0.2em]
 \epsilon_a e^{i\phi_a} &=& 
 \epsilon_{KM} \, 
 \frac{A_u(1/2,0)+A_u(1/2,1)+A_u(3/2,1)}
      {A_c(1/2,0)+A_c(1/2,1)+A_c(3/2,1)}
\ ,\\[0.2em] 
 \epsilon_{3/2} e^{i \phi} &=&
   \epsilon_{KM} \, 
 \frac{3 \, A_u(3/2,1)}
      {A_c(1/2,0)+A_c(1/2,1)+A_c(3/2,1)}
\ , \\[0.2em]
 \epsilon_T e^{i\phi_T} &=&
 \epsilon_{KM} \, 
 \frac{2 A_u(1/2,1)+2A_u(3/2,1)}
      {A_c(1/2,0)+A_c(1/2,1)+A_c(3/2,1)}
\ , \\[0.3em]
 q e^{i\omega} &=& 
  - \frac{1}{\epsilon_{KM}} \,
    \frac{A_c(3/2,1)}{A_u(3/2,1)}
\ , \\[0.3em]
 q_C e^{i\omega_C} &=&
  - \frac{1}{\epsilon_{KM}} \,
   \frac{A_c[1/2,1]+A_c[3/2,1]}{A_u[1/2,1]+A_u[3/2,1]}
\label{BBNSkpi}
\eeq
Thus, there are eleven independent isospin parameters for the $K \pi$ mode. 
At the moment, there is not enough data available to fix them all
independently. 

The factorizable part of the isopin amplitudes in the QCD factorization 
approach can be expressed  using the corresponding parameters 
$a_{i,\rm I}$ and $a_{i,\rm II}$ 
(the latter coefficients are restricted again to
the heavy quark limit) as given in \cite{Beneke:1999br}. 
\beq
\label{kpidecomposition}
A_c^F(3/2,1) &=& \frac{A_{\pi K}}{2} \left(
    - r_\chi^K \,  a_{8}^c 
    - a_{10}^c \right) + \frac{A_{K \pi}}{2} \left( - a_7 + a_9
\right) 
\ , \\[0.2em]
        A_u^F(3/2,1) &=& \frac{A_{\pi K}}{6} \left(
    2 a_{1} - 3  r_\chi^K \,  a_{8}^u 
    - 3 a_{10}^u \right) + \frac{A_{K \pi}}{6} \left( -2 a_2 + 3 a_7 - 3 a_9
\right) 
\ , \\[0.2em]             
A_c^F(1/2,1) &=& \frac{A_{\pi K}}{2} \left(
    - r_\chi^K \,  a_{8}^c 
    - a_{10}^c \right) + \frac{A_{K \pi}}{2} \left( a_7 - a_9
\right) 
\ , \\[0.2em]
A_u^F(1/2,1) &=& \frac{A_{\pi K}}{12} \left(
    - 2 a_{1} - 3  r_\chi^K \,  a_{8}^u 
    - 3 a_{10}^u \right) + \frac{A_{K \pi}}{6} \left( -2 a_2  - 3 a_7 + 3 a_9
\right) 
\ , \\[0.2em] 
A_c^F(1/2,0) &=& \frac{A_{\pi K}}{4} \left(4 a_4^c +  a_{10}^c 
    + 4  r_\chi^K \,  a_{6}^c  + r_\chi^K \,  a_{8}^c 
\right)
\ , \\[0.2em]
 A_u^F(1/2,0) &=& \frac{A_{\pi K}}{4} \left(
    2 a_{1} +   4 a_4^u +  a_{10}^u 
    + 4  r_\chi^K \,  a_{6}^u  + r_\chi^K \,  a_{8}^u \right)
\ .
\eeq

\subsection{$SU(3)_F$ connection between $B \to \pi\pi$ and 
$B \to \pi K$}

A well-defined procedure to reduce the number of independent
hadronic amplitudes in the analysis of
non-leptonic $B$\/ decays is to use the limit of
$SU(3)_F$ flavour symmetry.
For a comprehensive summary see Appendix~A
of \cite{Paz:2002ev}.\footnote{Note that we use
another  sign convention for the definition of isospin amplitudes}
Together with the structure of the effective hamiltonian, 
$SU(3)_F$ symmetry results in the following relation between the 
isospin amplitudes for $B \to \pi\pi$ and $B \to \pi K$ decays
\beq
A_p(2,3/2) &=&  A_p(3/2,1)
\ , \qquad p=u,c \ .
\label{firstrelation}
\eeq
These equations  represent in principle four additional 
relations between the real parameters entering the definition of the isospin 
amplitudes, namely two phase relations and two  ones connecting the moduli.
Using the parametrization (\ref{topol},\ref{xparam}) and
the approximate relation (\ref{FIERZ}) 
for $B \rightarrow \pi \pi$ amplitudes together with the parametrization 
for $B \rightarrow K \pi$ amplitudes quoted above, Eqs.~(\ref{firstrelation})
translate into
\beq
 q \, e^{i \omega} &\simeq& \delta_{\rm EW} 
 = - \frac{3}{2\epsilon_{KM}} \, \frac{C_9+C_{10}}{C_1+C_2} \simeq 0.69 \ , 
\label{firstsu3a}
\eeq
and
 \beq
  \frac{|\epsilon_{3/2} \, P|}{|\lambda_u^{(s)}|} &=&
  \frac{|(1+x \, e^{i\Delta}) \, \tilde T |}{|\lambda_u^{(d)}|} \ . 
\label{firstsu3} 
\eeq 
The first  equality states that in the $SU(3)_F$ limit
the parameter $q \, e^{i\omega} $ is a pure short-distance quantity, i.e.\
a real number that can be calculated in terms 
of Wilson coefficients and CKM factors. This is a well-known result
where only the structure of the electroweak effective hamiltonian 
enters \cite{Fleischer:1995cg,Neubert:1998pt}.  
The second relation connects the overall normalization, i.e.\
the dominant amplitudes $\tilde T$ for $B \to \pi\pi$, 
and $P$ for $B \to \pi K$ in the
naive factorization approach.
These are known
to receive sizeable $SU(3)_F$ breaking corrections, already in
naive factorization through the
ratio
\beq
  \frac{A_{\pi K}}{A_{\pi\pi}}=
  \frac{F^{B \to \pi}(m_K^2) f_K}{F^{B \to \pi}(m_\pi^2) f_\pi} \simeq 
  \frac{A_{K\pi}}{A_{\pi\pi}}  
 =\frac{F^{B \to K}(m_\pi^2) }{F^{B \to \pi}(m_\pi^2)}
&\simeq& 1.2 - 1.3  \ .
\label{eq:naivSU3}
\eeq

There is no further {\it exact} $SU(3)_F$ connection 
between $B \rightarrow \pi \pi$ and 
$B \rightarrow K \pi$ decays, because the $SU(3)_F$ multiplets involve also
other decays, namely $B \rightarrow K K$ and 
 $B_s \rightarrow K \pi$ decays for $\Delta S =0$ 
and $B_s \rightarrow K K$ and $B_s \rightarrow \pi \pi$ decays for 
$\Delta S = 1$.
However, using additional information by anticipating some experimental
data, one can derive further $SU(3)_F$ relations between the $B \to \pi \pi$
and $B \to K \pi$ modes.

If one {\it assumes} that 
\mbox{$A[B^0 \to K^+K^-] \to 0$} (which 
in the $SU(3)_F$ limit corresponds  to
\mbox{$A[B^0_s \to \pi^+\pi^-] \to 0$})  $SU(3)_F$ symmetry implies 
the following additional relations between isospin amplitudes 
\beq
A_p(0,1/2) &=& 
 A_p(1/2,0) -A_p(1/2,1) \ ,  \qquad p=u,c
\label{secondrelation}
\eeq
In the factorization approach this corresponds to neglecting
exchange/annihilation topologies.
It can be easily translated into four additional relations 
conecting two phases and two moduli in 
the $\pi\pi$ and in the $K \pi$ mode.  

Often some further empirical 
assumptions are used  within  the $SU(3)_F$ analysis, namely  
${\cal A}_{\rm dir}[B^\pm \to \pi^\pm K^0] = 0 $ or
$BR[B^\pm \to K^+K^0] = 0$. This allows to
eliminate two more parameters within  the $SU(3)_F$ analysis, namely 
(see (\ref{BBNSkpi})) 
\beq
\epsilon_a \, e^{i\phi_a} = 0 \,. 
\label{thirdrelation}
\eeq

Clearly, a complete  analysis of the systematic error 
of such a procedure has to take into account an estimate 
of  possible $SU(3)_F$ breaking effects,
and of the uncertainties related to the
phenomenological/empirical assumptions.
The simplest approach to estimate 
flavour-symmetry violating effects, is to restrict oneselves
to factorizable amplitudes as in (\ref{eq:naivSU3}).
However, in case of large 
non-factorizable QCD effects, which appear to be necessary to explain the 
$B \to \pi \pi$ data, one consequently has to take 
into account the possibility of sizeable flavour-symmetry 
violating effects for such contributions, too.
Naively factorizable $SU(3)_F$ breaking effects thus can only
provide an order-of-magnitude estimate
of flavour-symmetry breaking in general.

In the BBNS approach, apart from different
decay constants and form factors for pions and kaons,
$SU(3)_F$ breaking enters through the different (moments of)
light-cone distribution amplitudes. As we already pointed out
in Section~2, the universal treatment of non-factorizable
parameters $X_H$ and $X_A$ in {\it all}\/ $B \to PP$ isospin
amplitudes is questionable, because, on physical grounds, we
expect the related low-energy dynamics to depend on the
light quark masses in an essential way. In this context we
note that already in the framework of the BBNS approach, there
are additional sources of $SU(3)_F$ breaking.
According to \cite{Ball:1998je}, one has to consider strange-quark
mass corrections to the twist-3 amplitude $\phi_P^K$ 
(which in the BBNS approach parametrizes the endpoint
behaviour of the non-factorizable diagrams).
We have
(neglecting again all other terms that vanish in the asymptotic limit)
\beq
  \phi_p^{\pi,K}(u) &=& 1 + \rho_{\pi,K}^2 \left( -\frac52 \, C_2^{1/2}(2u-1)
                   - \frac{27}{20} \, C_4^{1/2}(2u-1) \right) 
\ ,
\eeq
where $C_n^m(\xi)$ are Gegenbauer polynomials, and 
\beq
   \rho_K^2 &=& \frac{m_s^2}{m_K^2} = {\cal O}(m_s/\Lambda) 
   \simeq 5-10\% \, \qquad \rho_\pi^2 \simeq 0 \ .
\eeq
In particular, at the endpoint, $u\to 1$ we obtain
\beq
 1-\frac{\phi_p^K(1)}{\phi_p^\pi(1)} &\simeq& 15 - 35 \%
\eeq
which is the typical size of flavor symmetry breaking 
usually assumed in other phenomenological analyses.
Notice that the above considerations 
point to {\it smaller}\/ non-factorizable effects for
kaons than for pions, which is different from the
behavior of the factorizable terms, where $f_K>f_\pi$
and $F^{B \to K} > F^{B \to \pi}$.
As this estimate does not 
include any dynamics related to the actual strong rescattering,
we cannot exclude that the true effect might be even
larger.

Finally, we note the parameter
$q_c e^{i\omega_c}$ in (\ref{BBNSkpi}) is neglected 
in many $SU(3)_F$ analyses (see for example \cite{Buras:2004ub}),
using the fact 
that in the limit of naive factorization, this parameter  
is {\it colour-suppressed}. 
However, also the parameter $x$ within the $B \rightarrow \pi\pi$  
is small in naive factorization due
to this argument, but nevertheless that parameter is found to be 
large when compared to experimental data (see section 2).
 
Thus, we conclude that in view of large strong 
phases and large contributions from
colour-suppressed   topologies observed in experiment,
any other prediction, following from
the factorization approximation, should be critically re-analysed.

\subsection{$R_n$-puzzle}

The present data on the $K \pi$ branching ratios can be expressed
in terms of three ratios  \cite{Charles:2004jd,Chao:2003ue,Aubert:2003sg}: 
\beq
R &=& \frac{\tau_{B^+}}{\tau_{B^0}}\,
    \frac{{\rm BR}[B_d^0 \to \pi^-K^+]+{\rm BR}[\bar B_d^0 \to \pi^+K^-]}
        {{\rm BR}[B_d^+ \to \pi^+K^0]+{\rm BR}[B_d^- \to \pi^- \bar K^0]}
\;=\; 0.91^{\,+0.08}_{\,-0.07}
\ ,\\[0.3em]
 R_n &=& \frac12 \, 
    \frac{{\rm BR}[B_d^0 \to \pi^-K^+]+{\rm BR}[\bar B_d^0 \to \pi^+K^-]}
        {{\rm BR}[B_d^0 \to \pi^0K^0]+{\rm BR}[\bar B_d^0 \to \pi^0 \bar K^0]}
\;=\; 0.78^{\,+0.11}_{\,-0.09}
\ ,\\[0.3em]
 R_c  &=& 2 \, 
    \frac{{\rm BR}[B_d^+ \to \pi^0 K^+]+{\rm BR}[B_d^- \to \pi^0 K^-]}
        {{\rm BR}[B_d^+ \to \pi^+ K^0]+{\rm BR}[B_d^- \to \pi^- \bar K^0]}
 \;=\; 1.16^{\,+0.13}_{\,-0.11}
\ .
\eeq
This result appears somewhat anomalous,
when compared, for example, with the approxi\-mate sum rule 
proposed in \cite{Lipkin:1998ie,Gronau:1998ep,Matias:2001ch}            
which leads to the prediction  $R_c = R_n$, or from  
the comparison with the  $\pi\pi$ data using the $SU(3)_F$ 
symmetry approach. In particular, the ratio $R_n$ appears to
be smaller by about two standard deviations than one would
expect. It is important to note that 
the deviation of $R_n$ and $R_c$ from unity is solely due to isospin-breaking
effects \cite{Matias:2001ch}. 
The amount of short-distance isospin breaking in the
Standard Model is too small to explain the experimental number.
Whereas the authors of  \cite{Buras:2004ub} 
argue that this may point
to an interesting avenue towards new physics in electroweak
penguin operators, the collaboration in \cite{Charles:2004jd}
considers this deviation as a statistical fluctuation, which
is consistent with the Standard Model -- 
even when the {\it generalized}\/ $SU(3)_F$ constraints,
(\ref{firstrelation}) and  (\ref{secondrelation}), are enforced.  
Not surprisingly, the $B \to K \pi$ data has 
triggered several new-physics analyses 
(for the very recent literature see 
\cite{Buras:2004ub,Nandi:2004dx,Xiao:2004zt,Khalil:2004yb,Mishima:2004um}).

Apart from the theoretical questions about the interpretation of the
data, there are also two experimental issues 
which have to be clarified:
\begin{itemize}
\item Radiative corrections to decays with charged particles in the final
states may not have been taken into account properly
in the experimental analysis, an effect which is expected to lead 
to an increased branching ratio of these modes \cite{Charles:2004jd}
and which could bring $R_n$ closer to unity. 
\item 
It has also been argued in \cite{Gronau:2003kj}
that the present pattern 
could result from underestimating the $\pi^0$ detection efficiency 
which implies an overestimate for any branching ratio involving a $\pi^0$.
The authors of \cite{Gronau:2003kj}
propose therefore to consider the 
ratio $(R_n R_c)^{1/2}$ in which the $\pi^0$ detection efficiency
cancels out. Again, the experimental value for this quantity 
is closer to unity.
\end{itemize}

But even when the experimental situation is clarified and the experimental 
accuracy will be significantly improved, one has to critically 
analyse whether the data on $B \to \pi\pi$ and $B \to \pi K$ decays 
point to new physics 
(for example to isospin breaking via new 
degrees of freedom as discussed in \cite{Fleischer:1997ng,Grossman:1999av}), 
or whether they can be explained by non-factorizable 
$SU(3)_F$- or isospin-violating QCD and QED effects within the SM.
Obviously, the inclusion of more hadronic parameters to include
such effects can only improve the phenomenological situation, compared
to the analyses based on flavour-symmetry or the constrained BBNS
scenario. On the other hand, the order-of-magnitude of 
flavour-symmetry breaking effects should be consistent with
the theoretical and experimental estimates of 
the non-factorizable contributions in  the $B \to \pi\pi$ sector.

In order to illustrate this point, we discuss a 
toy model which is inspired by the result of the  previous section.  
We start with the assumption that  
all nonfactorizable contributions are saturated by 
long-distance QCD and QED penguin contributions.
To estimate the possible numerical effect, we multiply
the factorizable QCD and QED penguin by 
a common enhancement factor  $(1 - \Delta P_ {\rm NF})$,
where $\Delta P_{\rm NF}$ is a complex parameter.
Within the BBNS approach (see (\ref{AF}) and (\ref{kpidecomposition})) 
the QCD penguin function $P^{QCD}_p$  
appears  in the combination $a_4^{p}+ a_6^{p}\, r_{\chi} $, 
while the penguin 
function  $P^{EW}_p$ occurs in the combination  
$a_{10}^p  + a_8^{p}  \,r_{\chi}$. The two $SU(3)_F$ constraints, 
(\ref{firstrelation}) and (\ref{secondrelation}),
are  both compatible with this procedure by construction.  
From the $B \to\pi\pi$ decays, we find that the long-distance 
QCD penguins require an enhancement by about an order of magnitude
(long-distance QED penguins are negligible in $B \to \pi\pi$).
Consequently, the long-distance QED effects in $B \to \pi K$ 
are enhanced by the same amount, which leads
to significant changes in the
isospin-violating parameters $q e^{i\omega}$ and
$q_C e^{i\omega_C}$. 
We find typical values in the 
range $0.3 < q < 0.8$, 
$-30^\circ < \omega < 30^\circ$, 
$0.05 < q_C < 0.3$, and   $-150^\circ < \omega_C < 150^\circ$.

More generally, 
in our framework the expansion parameter that suppresses
long-distance isospin-violating effects
in $B \to \pi K$ is given by
 $$ r_c(0,1/2) \, \cdot \, \frac{\alpha_{\rm em}}{\epsilon_{KM}}
          \ \simeq \ 5-10\% \ ,
 $$
where the latter number refers to typical numbers for
$r_c(0,1/2)=0.1-0.2$ observed in $B \to \pi\pi$.
This seems to be in the right ball park to at least partly explain
the deviation of $R_n$ from unity.

If realized in nature, the latter scenario should also
have some impact for other ``puzzling'' $B$ decays: 
In $B \to \phi K_S$ the apparent hierarchy 
$r_u(0,1/2) > r_c(0,1/2)$ in $B \to \pi\pi$ decays
may translate into a moderate enhancement of the CKM-suppressed
penguin amplitude. This may lead to some deviation of the
extracted value of $\sin2\beta_{\rm eff}$ from the value
found in $B \to J/\psi K_S$, but is probably not sufficient
to explain the present BELLE measurement \cite{Abe:2003yt}
for this quantity.
Non-factorizable penguin contributions can be even more
important in $B \to \eta'K$ decays, because for decays into singlet
mesons some non-factorizable operators appear at lower
order in the $1/m_b$ expansion than for the octet ones
(see the discussion in Section~\ref{sec:scet}).
According to \cite{Beneke:2002jn} this mainly
implies that the theoretical uncertainties in such decays presently are
too large to extract information on short-distance
quantities in a reliable way.

\vspace{2em}

Our main conclusion is a conservative one, namely that the 
unsatisfactory theoretical understanding of non-factorizable 
power-suppressed effects in charmless non-leptonic $B$\/ decays prevent
us from identifying new-physics effects in these observables,
at least on the level of present experimental significance of
certain ``puzzles''. On the other hand, the comparison of 
different possible approximation schemes, used to reduce the number
of unknown hadronic parameters, gives a handle to estimate
model-dependent systematic effects in CKM studies.
It may also shed some light on the
dynamical origin of non-factorizable effects, which
may stimulate further studies with non-perturbative methods.
In particular, a systematic classification of power-suppressed
matrix elements from non-factorizable SCET operators should
give an alternative scheme compared to the traditional
classification in terms of flavour topologies.

\section*{Acknowledgements}

We thank Ahmed Ali, Martin Beneke, Robert Fleischer, Joaquim Matias, 
Salim Safir, Lalit Sehgal, and Iain Stewart for interesting 
and helpful discussions. T.H. also thanks  Raul Hernandez Montoya 
for his kind hospitality at the University
of Veracruz where the present work was finalized.

\end{document}